\documentclass{aa}  
\usepackage{graphicx}
\usepackage{txfonts}
\usepackage{comment}
\usepackage{adjustbox}

\usepackage{xspace}
\newcommand{\kms}{km~s$^{-1}$\xspace}

\newcommand{\Msunyr}{M$_{\odot}$~yr$^{-1}$\xspace}

\newcommand{ \ha} {H$\alpha$\xspace}
\newcommand{ \OIII} {[O\,\textsc{iii}]\xspace}

\usepackage{multirow}
\usepackage{booktabs}

\newcommand{\prism} {PRISM\xspace}

\usepackage[breaklinks=true, colorlinks=true, citecolor=blue, linkcolor=blue,urlcolor=blue]{hyperref}

\usepackage{threeparttable}

\begin{document}

\title{GA-NIFS: The interplay between feedback and star formation at $3~<~z~<~9$ probed by JWST/NIRSpec IFU}

\author{
L. Ulivi\inst{\ref{iCAB}}\thanks{e-mail: lulivi@cab.inta-csic.es}
\and M. Perna \inst{\ref{iCAB}}
\and S. Arribas \inst{\ref{iCAB}}
\and B.~Rodr\'iguez~Del~Pino\inst{\ref{iCAB}}
\and P.~G.~Pérez-González\inst{\ref{iCAB}}
\and I.~Lamperti \inst{\ref{iCAB}}
\and C.~Marconcini\inst{\ref{iOAA}}
\and H.~\"{U}bler\inst{\ref{iMPE}}
\and F.~D'Eugenio\inst{\ref{iKav},\ref{iCav}}
\and T.~B{\"o}ker\inst{\ref{iESAusa}}
\and A.~J. Bunker\inst{\ref{iOxf}}
\and S.~Carniani \inst{\ref{iNorm}}
\and S.~Charlot\inst{\ref{iSor}}
\and R~Maiolino\inst{\ref{iKav},\ref{iCav}, \ref{iUCL}}
\and P. Alvarez \inst{\ref{iCAB}}
\and E.~Bertola\inst{\ref{iOAA}}
\and G.~Cresci\inst{\ref{iOAA}}
\and M.~Curti\inst{\ref{OAB}}
\and M.~Hamed\inst{\ref{iCAB}}
\and L.~R. Ivey\inst{\ref{iKav},\ref{iCav}}
\and G.~C.~Jones\inst{\ref{iOxf},\ref{iKav},\ref{iCav}}
\and E.~Parlanti\inst{\ref{iNorm}}
\and R.~Pascalau\inst{\ref{iKav},\ref{iCav}}
\and C. Prieto Jimenez \inst{\ref{iCAB}}
\and B.~Trefoloni\inst{\ref{iNorm}}
\and G.~Venturi\inst{\ref{iNorm}}
\and S.~Zamora\inst{\ref{iNorm}}
%
}
\institute{
Centro de Astrobiolog\'ia (CAB), CSIC--INTA, Cra. de Ajalvir Km.~4, 28850 -- Torrej\'on de Ardoz, Madrid, Spain\label{iCAB}
\and
INAF - Osservatorio Astrofisico di Arcetri, Largo E. Fermi 5, I-50125 Firenze, Italy\label{iOAA}
\and
Max-Planck-Institut f\"ur extraterrestrische Physik (MPE), Gie{\ss}enbachstra{\ss}e 1, 85748 Garching, Germany\label{iMPE}
\and 
Kavli Institute for Cosmology, University of Cambridge, Madingley Road, Cambridge, CB3 0HA, UK\label{iKav}
\and
Cavendish Laboratory - Astrophysics Group, University of Cambridge, 19 JJ Thomson Avenue, Cambridge, CB3 0HE, UK\label{iCav}
\and
European Space Agency, c/o STScI, 3700 San Martin Drive, Baltimore, MD 21218, USA\label{iESAusa}
\and
Department of Physics, University of Oxford, Denys Wilkinson Building, Keble Road, Oxford OX1 3RH, UK\label{iOxf}
\and
Scuola Normale Superiore, Piazza dei Cavalieri 7, I-56126 Pisa, Italy\label{iNorm}
\and
Sorbonne Universit\'e, CNRS, UMR 7095, Institut d’Astrophysique de Paris, 98 bis bd Arago, 75014 Paris, France\label{iSor} 
\and
Department of Physics and Astronomy, University College London, Gower Street, London WC1E 6BT, UK\label{iUCL}
\and
INAF - Osservatorio di Astrofisica e Scienza dello Spazio, Via Piero Gobetti 93/3, 40129, Bologna, Italy\label{OAB}
}
    \titlerunning{ Prism R100 GA-NIFS z>3}
    \authorrunning{Ulivi, L., et al.}


 
  \abstract
   {The study of starburst- and AGN-driven feedback is fundamental for understanding the processes that shape galaxy growth, quench star formation, and drive the co-evolution of galaxies and their central black holes. 
   We present a spatially resolved study of six galaxies at $3 < z < 9$, including starburst- and AGN-dominated systems, observed with \textit{JWST}/NIRSpec IFU in low- ($R \sim 100$) and high-resolution ($R \sim 2700$)  mode. Previous  analysis of $R \sim 2700$ data revealed ionized outflows in all these galaxies. We explore possible links between outflows and stellar population properties to assess the mechanisms driving galaxy evolution at early epochs. 
   We derive stellar masses and star formation histories (SFHs) using the spectral energy distribution fitting code \texttt{Prospector}, applied on both single spaxels and spatially integrated spectra, exploring parametric and non-parametric SFHs. 
   We find that unresolved stellar mass measurements are systematically underestimated, with discrepancies increasing toward higher sSFRs (up to $\sim 0.75$~dex), consistent with the so-called outshining effect. 
 Combining the spatially resolved stellar population analysis with the outflow properties, we find that the most massive galaxies $(M_\star>5\times10^{10}M_\odot)$ with strong AGN-driven ionized outflows show evidence of past quenching episodes occurring within the last $\sim 100$–$300$ Myr, mainly in the nuclear regions ($r$ < 3~kpc). In contrast, higher-redshift ($z>5$) and less massive $(M_\star<10^{10}M_\odot)$ starburst galaxies with powerful (starburst-driven) outflows appear to have experienced continuous growth, with no clear sign of quenching. 
 Massive galaxies with weaker outflows do not show evidence of quenching in their history. One massive AGN host (GS20936) shows evidence for a recent (10--30~Myr) rejuvenation phase, likely fueled by a recent major merger. This interpretation is supported by TNG300 cosmological simulations, which indicate that  $\sim 5\%$ of galaxies with $M_\star > 5 \times 10^{10}\, \mathrm{M}_\odot$  undergo similar rejuvenation events driven by merging at $z\sim 3$. 
These results suggest that (AGN-driven) outflows can already play a key role in shaping the SFHs of massive galaxies at early cosmic epochs, while also highlighting the importance of spatially resolved analyses for accurately reconstructing their evolutionary pathways.
  
   }



    \maketitle
%

\section{Introduction}

Understanding how galaxies assemble their stellar mass ($M_{\star}$) and cease their star formation is a central problem in galaxy evolution. A key observational framework is the tight correlation between $M_{\star}$ and the star formation rate (SFR), commonly referred to as the star-forming main sequence \citep[e.g.,][]{Brinchmann2004,Daddi2007,Whitaker2014}. The relatively small scatter ($\sim$ 0.3 dex) of this relation suggests that most star-forming galaxies evolve in a quasi steady-state regime, where gas accretion, star formation, and feedback are self-regulated. In this context, galaxies are expected to oscillate around the main sequence, undergoing alternating phases of enhanced (starburst-like) and suppressed (quiescent) star formation. 

Stellar masses and SFR are fundamental properties of galaxies, but accurately measuring them from integrated spectra or integrated photometry could lead to large uncertainties or bias towards young stellar populations that contribute strongly to the luminosity but little to the total stellar mass. Spatially resolved studies are essential for minimizing systematic errors inherent in integrated measurements, allowing for a more accurate treatment of dust attenuation, a clearer separation of mixed stellar populations across different regions, and consequently more accurate spectral energy distribution (SED) fitting, leading to more reliable estimates of physical parameters. Moreover, different regions within a galaxy often exhibit distinct assembly histories, providing insights into how galaxies grow and quench and, in general, a more comprehensive picture of a galaxy’s evolution (\citealt{Tacchella2015,Nelson2016,Belfiore2017, Coenda2019, Abdurrouf2018, Abdurrouf2023}). 

In the local Universe, spatially resolved observations have established that star-forming galaxies generally assemble their stellar mass in an inside-out fashion \citep{MunozMateos2007,Pezzulli2015,Frankel2019,Bluck2020}. This growth pattern is often accompanied by a progressive suppression of star formation from central regions outward, which may be either temporary or permanent. More in general, several mechanisms have been proposed to drive quenching, including 
internal processes such as feedback from the active galactic nucleus (AGN)
 (e.g \citealt{Fabian2012, King2015, Marconcini2025_nature}), and external processes such as environmental effects (e.g. ram-pressure stripping and strangulation; e.g. \citealt{McGee2014, Whitaker2021, Williams2021}). These mechanisms can operate on different spatial and temporal scales, leading to diverse quenching pathways that depend on galaxy mass and environment \citep{Schaefer2017,GonzalezDelgado2017,Medling2018,Spindler2018}. Disentangling the relative importance of these processes remains challenging even at low redshift, and becomes increasingly difficult toward higher redshift, where observational constraints are more limiting.


Understanding the mechanisms regulating star formation at high redshift is particularly important because galaxies observed only a few billion years after the Big Bang provide direct constraints on the timescales and physical processes driving early galaxy assembly and quenching. In recent years, evidence has emerged for galaxies experiencing quenching already well before cosmic noon
(e.g. \citealt{Carnall2022,Nanayakkara2024,deGraaff2025}).
At $z\gtrsim 3$, the Universe is younger than $\sim$2 Gyr, which places strong limits on the timescales over which quenching can occur. 
This suggests that quenching at high redshift must involve fast and efficient mechanisms, potentially linked to intense feedback from extreme early star formation and AGN activity, as also suggested from recent theoretical studies.\citep{Donnari2019, Tacchella2020,Tacchella2022,Dome2024,Lagos2024, Xie2024}.

Thanks to its unprecedented sensitivity and spatial resolution in probing the rest-frame UV-to-optical emission of distant galaxies, the \textit{James Webb Space Telescope} (\textit{JWST}, \citealt{Gardner2023}) is now providing unprecedented constraints on both the onset (e.g. \citealt{Endsley2023,Cameron2023,Rinaldi2023,Caputi2024,Simmonds2024,Jones2026}) and quenching (e.g. \citealt{Nanayakkara2022,Marchesini2023,Looser2024,Turner2025,Yang2026}) of star formation at $z>3$.
However, the lack of direct constraints on spatially resolved SFHs still prevents a physical understanding of quenching, including its drivers, spatial progression, connection to outflows, duration, and possible rejuvenation. 

Rejuvenated galaxies, namely galaxies that have resumed their star formation after a quiescent phase, have been identified in several studies at $z < 2$ (\citealt{Treu2005,Chauke2019,Mancini2019,Belli2021, Cleland2021,Tacchella2022,Paspaliaris2023}) but at higher redshift have yet to be observed.
Several studies have suggested that rejuvenation may be linked to external processes such as mergers or gas inflow \citep{Kaviraj2009,Rathore2022,Tanaka2024}, 
but its overall role in galaxy evolution remains uncertain. 



In this work, we take advantage of spatially resolved \textit{JWST}/NIRSpec spectroscopic observations to investigate the interplay between star formation and outflows in six galaxies at $3 < z< 9$, including both starburst- and AGN-dominated systems. By combining integrated and spaxel-by-spaxel analyses, we derive both global and spatially resolved SFHs, enabling us to assess how unresolved measurements may bias the inferred stellar population properties (e.g., due to the outshining effect \citealt{Hamed2026}). Furthermore, we examine the spatial distribution of SFHs in relation to the spatial location of outflowing ionized gas, to understand the role of feedback in regulating star formation and driving quenching at early cosmic times. 
Finally, we exploit the TNG300 simulations to place our results in a broader evolutionary context, investigating whether quenching can be followed by a subsequent rejuvenation phase, its contribution to galaxy growth, and how frequently it may occur.

This paper is structured as follows. In Sect.~\ref{sec:Obs} and Sect.~\ref{sec:Sample}, we present the observations and describe the selected sample. In Sect.~\ref{sec:Analysis}, we outline our analysis framework and derive both integrated and spatially resolved physical properties. In Sect.~\ref{sec:Results}, we present and discuss our results, and in Sect.~\ref{sec:Conclusions} we summarise our conclusions. Throughout this work, we assume $ \Omega_{\rm m}=0.309$ and $H_0=67.66$ \kms Mpc$^{-1}$ \citep{Bennett2014}.

\section{Observations and data reduction}
\label{sec:Obs}

The data analysed in this work were obtained with the \textit{JWST} Near-Infrared Spectrograph (NIRSpec) Integral Field Unit (IFU; \citealt{Jakobsen2022,Boker2022, Rigby2023}) as part of the Galaxy Assembly with NIRSpec Integral Field Spectroscopy (GA-NIFS\footnote{\url{https://ga-nifs.github.io}}) Guaranteed Time Observations (GTO) program (
see \citealt{Perna2023a} for an early overview). 
The survey targets 55 AGN and star-forming galaxies over the redshift range $z \sim 2-11$, enabling spatially resolved studies of galaxy structure, star formation, and feedback processes.

The raw data were retrieved from the Barbara A. Mikulski Archive for Space Telescopes (MAST) and processed using the \textit{JWST} Science Calibration Pipeline (v1.20.2) with Calibration Reference Data System (CRDS) context {\it jwst\_1464.pmap}. 
In addition to the default pipeline, we applied a set of custom procedures to improve data quality. These include correction of $1/f$ (pink) noise, and masking of residual cosmic rays (including ``snowballs'') and signal from failed open microshutter array shutters. A detailed description of these steps is provided in \citet{Perna2025_ring}. 
We adopted the standard outlier rejection algorithm and resampled the data to a spatial scale of $0.05''$ per spaxel using the drizzle algorithm.

We further corrected the noise estimates in the final cubes, as the pipeline-provided uncertainties were found to underestimate the true noise level. Following \citet{Ubler2023}, we rescaled the error spectra in each spaxel based on the empirical noise measured in line-free spectral regions.

\begin{figure*}
    \centering
    \includegraphics[width=1\linewidth]{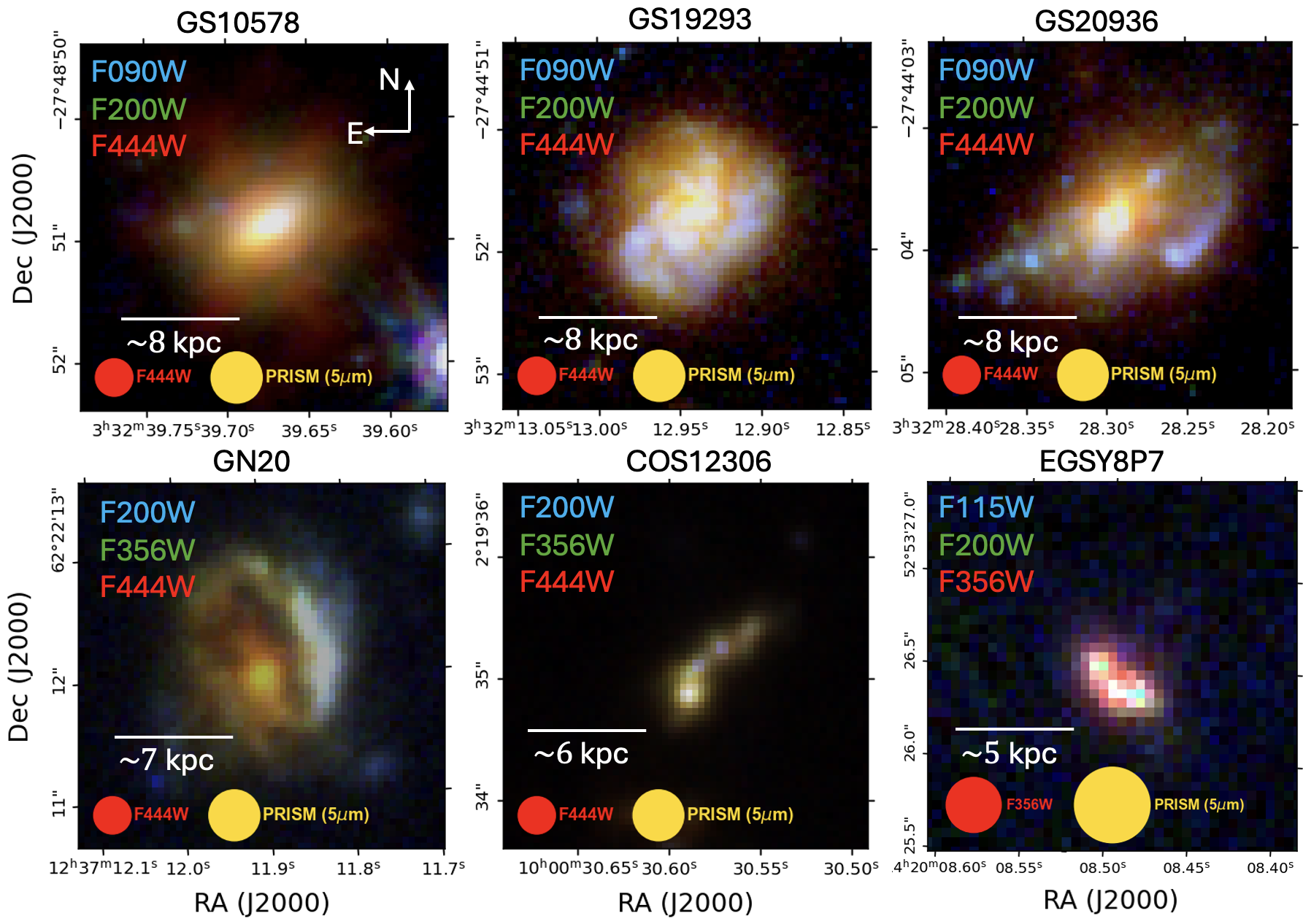}
    \caption{NIRCam three colours images of our sample. For GS10578, GS19293, and GS20936 we use the F090W, F200W, and F444W filters. For GN20 and COS12306, F356W is used in place of F090W. For EGSY8p7, the highest-redshift source in our sample, we adopt the F115W, F200W, and F356W filters. The red circles represent the largest PSF FWHM used for the RGB image, while the gold circles represent the R100 PSF at 5~$\mu$m. North is up and East is to the left. White horizontal lines represents 1". }
    \label{fig:three_colors}
\end{figure*}

\section{Sample selection}\label{sec:Sample}

\begin{table*}
\tabcolsep 5pt 
\centering
\caption{Description of the NIRSpec IFU R100 observations, and sample properties.
}
\label{tab:targets}
\begin{tabular}{lcccccccccc}
\hline\hline
Target & Program ID & RA & Dec & $z$ & $t_{\rm exp}$ &Spatial Sampling & $\lambda_{\rm SED}^{rest}$  & L$_{\mathrm{AGN}}$ & $\Delta$MS  \\
&         & [deg] & [deg] &     & [s] & [pc] & [$\mu$m] & [$10^{45}$erg/s] & [dex] & \\

\hline
GS10578 & 1216 & 53.16531 & $-27.81413$ & 3.064 & 3968  & 390 & 0.18--1.10 & 6.5$\pm$0.7 & <--0.35 \\
GS19293 & 1216 & 53.05393 & $-27.74771$ & 3.119  & 3968 & 390 & 0.20--1.25  & <1.1 &  <0.55  \\
GS20936 & 1216 & 53.11790 & $-27.73438$ & 3.243 & 3968 & 390 & 0.18--1.18 & 5.3 $\pm$0.5 & 0.45  \\
GN20 & 1264 & 189.29958 & $+62.37003$ & 4.055 & 3560 & 360 & 0.19--1.02 & <40 & 0.57 \\
COS12306 & 1217 & 150.12746 & $+2.32643$ & 5.843 & 3968 & 300 & 0.15--0.74  & -- & 0.64 \\
EGSY8p7 & 1262 & 215.03542 & $+52.89072$ & 8.683 & 3968 & 240 & 0.13--0.52 &  -- & 0.54 \\
\hline
\end{tabular}

\tablefoot{Spatial Sampling corresponds to 0.05" per pixel. $\lambda_{\rm SED}^{\rm rest}$ represents the rest-frame wavelength interval used in the SED fitting procedure. L$_{\mathrm{AGN}}$ are taken from \cite{Venturi2025} for GS sources, and from \cite{Riechers2014} for GN20. $\Delta$MS represents the offset (in dex) from the star-forming main sequence at the target redshift as defined by \cite{Clarke2024}.}

\end{table*}


Our selection is designed to enable a robust analysis of stellar population properties on resolved scales, focusing specifically on galaxies classified as AGN or starbursts with evidence of ionized outflows, thereby allowing a direct comparison between regions strongly influenced by the outflows and those that are unaffected.. 
To this end, from the GA-NIFS parent sample, we selected a subsample of galaxies at $3\lesssim z\lesssim 9$ satisfying the following criteria. First, the galaxies must have both low- ($R\sim100$) and  high ($R\sim2700$) or medium-spectral resolution ($R\sim1000$) \textit{JWST}/NIRSpec IFU observations, ensuring reliable constraints on both the stellar continuum and the emission-line kinematics. Second, targets were required to be either AGN-dominated systems, based on X-ray and optical diagnostic line ratio diagrams classifications (\citealt{Perna2025, Ubler2024}), or starburst systems. 
Following \citet{RodriguezDelPino2026}, starbursts were defined as galaxies lying more than 3 times above the star-forming main sequence at the corresponding redshift, adopting the relation from \citet{Clarke2024}. 

Third, the continuum emission had to be spatially resolved with sufficient signal-to-noise ratio in each spaxel (S/N $>3$, see Sect.~\ref{sec:apertureforintegratedspectra}) to allow for a reliable analysis of the stellar populations; galaxies with spatially unresolved or undetected continuum emission were therefore excluded. Finally, all selected systems were required to exhibit ionized outflows traced by broad (i.e. $\sigma >250$~km/s) components of \OIII and H$\alpha$ emission lines, enabling a direct comparison between outflow properties and the spatially resolved stellar populations (\citealt{Venturi2025,RodriguezDelPino2026}).

Out of the 55 systems in the GA-NIFS sample, 22 have observations with the R100 grating. Ten of these are inactive MS or starburst galaxies with weak or absent ionized outflows, and are therefore not considered further; a parallel study will focus exclusively on these systems (Ulivi et al., in prep.). Of the remaining 12 systems, six were excluded for the following reasons.
In the Jekyll \& Hyde system (\citealt{Glazebrook2017,Perez-Gonazalez2025}), the former galaxy does not show evidence of the presence of AGN  while Hyde is spatially unresolved in the R100 data.
SPT0311–58 
was excluded because no clear signatures of ionized outflows are detected \citep{Arribas2024}. Himiko consists of three compact sub-systems with projected separations of $\sim3-4$~kpc, 
preventing the derivation of resolved stellar populations (\citealt{Kiyota2025}). COS30 and MACS1149-JD1 were excluded because the continuum S/N falls below our threshold for spatially resolved analysis; we refer the reader to \citet{Scholtz2026} and \citet{Marconcini2024}, respectively, for more detailed studies of these sources.

The final sample comprises six galaxies: GS10578, GS19293, GS20936, GN20, EGSY8p7, and COS12306.  These sources span a wide range of redshifts ($3<z<9$) and physical conditions, including quiescent systems hosting luminous AGN and starburst galaxies.  Table~\ref{tab:targets} summarizes the main properties of the targets, including AGN bolometric luminosities and MS offset, and an overview of the R100 observation details. While this sample is limited in size, the high quality of the spatially resolved JWST/NIRSpec IFU spectra allow us to gain insight into the physical processes shaping galaxy growth, feedback, quenching, and possible rejuvenation in the early Universe. Figure~\ref{fig:three_colors} shows NIRCam three-colour images of the sample, obtained by JADES (\citealt{Eisenstein2026}). 
In the following, we briefly describe the main properties of each target.

\paragraph{GS10578}
is a massive (M$_\star = 2\times10^{11}$~M$_\odot$) galaxy at $z = 3.064$, classified as quiescent (SFR $<40$~\Msunyr) despite hosting a luminous AGN (\citealt{DEugenio2024}). The AGN is heavily obscured ($N_{\rm H} \sim 5.5 \times 10^{23}\ \mathrm{cm^{-2}}$; \citealp{Circosta2019}) and identified through optical emission-line diagnostics 
and strong X-ray emission ($L_{\rm 2-10~keV} \sim 8 \times 10^{44}\ \mathrm{erg\ s^{-1}}$; \citealp{Luo2017_cat}). NIRSpec IFU observations reveal powerful atomic ionized and neutral outflowing gas \citep{DEugenio2024,Venturi2025}, while its cold molecular gas content places the most stringent upper limits on the gas mass for a quiescent galaxy at high-z ($<10^{9.1}$~M$_\odot$, corresponding to $<0.8\%$ of $M_\star$), suggesting that gas inflows have been balanced by AGN feedback during its evolution, either ejecting material or preventing its accretion onto the galaxy \citep{Scholtz2026}.
GS10578 resides in a complex environment. NIRSpec data reveal an additional AGN separated by $\sim$5~kpc from the nucleus of GS10578, while VLT/MUSE observations identify an extended Ly$\alpha$ nebula and a luminous Ly$\alpha$ emitter at a projected distance of $\sim$~30 kpc (\citealt{Perna2025}). New JWST observations confirm the AGN nature of this third source (\citealt{Perna2026lae2}). 

{\noindent \normalsize\sffamily GS19293}
($z = 3.119$) was initially targeted  for the GA-NIFS survey as a star-forming galaxy. Although X-ray undetected in the \textit{Chandra} Deep Field South 
(7~Ms; \citealp{Luo2017_cat}), optical emission-line diagnostics revealed the presence of an AGN \citep{Perna2025}. SED fitting performed by Circosta et al. (in prep.) yields a stellar mass of $2\times10^{10}$~M$_\odot$ and a $1\sigma$ upper limit on the SFR of 77~\Msunyr. 
Spectral analysis from \cite{Venturi2025} indicates a mostly blueshifted outflow with $\sigma \sim 200-300$~\kms, together with a smoothly rotating ionized gas disc with velocity dispersions $\lesssim 50\ \mathrm{km\,s^{-1}}$. No companions or tidal features are identified for this source (Fig.~\ref{fig:three_colors}), consistent with its regular rotating disk kinematics.

{\noindent \normalsize\sffamily GS20936}
($z = 3.243$) was selected as a distant red galaxy \citep{Wuyts2009}. It has an absorption-corrected X-ray luminosity of $L_{\rm 2-10~keV} \sim 1.3 \times 10^{43}\ \mathrm{erg\,s^{-1}}$ (\citealt{Luo2017_cat}). Optical emission-line diagnostics reveal the presence of an AGN \citep{Perna2025}. 
UV-optical to FIR photometry SED fitting yields M$_\star =10^{10}$~M$_\odot$ and SFR $= 41$~\Msunyr (Circosta et al. in prep).
As shown in Fig. \ref{fig:three_colors}, GS20936 is very extended, showing a complex morphology, with clumpy, asymmetric and disturbed structures indicative of an ongoing merger (see also Sect.~\ref{sec:rejuvenation}). Spectral analysis reveals \OIII and \ha outflow components ($\sigma > 250$~km s$^{-1}$) in the nuclear regions (\citealt{Venturi2025}).

{\noindent \normalsize\sffamily GN20.}
is a barred starburst galaxy located within a proto-cluster environment at $z \simeq 4.055$ (\citealt{Boogaard2026}). It exhibits an infrared luminosity of $L_{\mathrm{IR}} \sim 10^{13}\ L_\odot$, corresponding to a SFR $\sim 1800-3000$~\Msunyr \citep{Pope2005}, and a stellar mass of $M_\star \sim 1-6\times10^{11}M_\odot$ \citep{Tan2014}. X-ray observations provide only an upper limit, with L$_{2-10~\mathrm{keV}}$ $< 1.8 \times 10^{42}\ \mathrm{erg\ s^{-1}}$) (\citealt{Riechers2014}). The presence of an AGN is supported by the heavily obscured nuclear region (see Fig.~\ref{fig:three_colors}) showing \ha broad lines ($\sigma > 400~$\kms, that may be attributed to either BLR or outflows), and by the `WHAN' diagnostic diagram, which combines the \ha equivalent width with the [N\,\textsc{ii}]$\lambda6583$/\ha line ratio, yielding values consistent with the Seyfert regime (\citealp{CidFernandes2010,Ubler2024}). Further evidence of the presence of an AGN is presented in Hamed (in prep.).

{\noindent \normalsize\sffamily COS12306}
was identified as one of the most luminous and extended star-forming galaxies at z = 6 in the CFHT Legacy Survey \citep{Willott2013}, with M$_\star = 10^{10}$~M$_\odot$. Rest-frame optical emission lines place the galaxy at $z \sim 5.843$, and the SFR derived from H$\alpha$ ($162$~\Msunyr) indicates that it is undergoing a starburst phase.  Figure \ref{fig:three_colors} shows that the galaxy has a complex morphology with several clumps aligned across $\sim6$~kpc, possibly all part of the same merging system \citep{RodriguezDelPino2026}.

{\noindent \normalsize\sffamily EGSY8p7}
($z = 8.683$) is the highest-redshift source in our sample. It has been proposed as a potential progenitor of massive high-redshift quasars based on the detection of a broad H$\beta$ component \citep{Larson2023,Marques-Chaves2024}, although this feature was instead recently associated with starburst-driven outflows \citep{Zamora2025}. With a stellar mass of $(5-74)\times10^{8}$~M$_\odot$ and a SFR of 49-97~\Msunyr, EGSY8p7 is the most compact (see Fig.~\ref{fig:three_colors}) galaxy of our sample and it is 
composed of three distinct stellar clumps,  likely involved  in a  merging event responsible  for the enhanced SF (\citealt{Larson2023}). An outflow is observed in the \OIII emission line, with $\sigma \sim 300$~\kms and an extension of $\sim 1$~kpc \citep{Zamora2025}.

\section{Analysis}\label{sec:Analysis}
In this section, we describe the methods adopted for the stellar population analysis, performed both on the integrated spectra of each galaxy and on a spatially resolved, spaxel-by-spaxel basis. The SED fitting was performed using the same configuration for both the resolved and unresolved analyses, ensuring that any differences in the derived properties can be attributed solely to the effects of spatial resolution. Before discussing the details of the SED modelling, we summarize the additional data processing steps required to ensure reliable measurements, including background subtraction, spatial resolution homogenization, and the construction of apertures for resolved analysis and for extracting integrated spectra.

\subsection{Background subtraction and PSF homogenization}

We performed the background subtraction directly on the datacubes by selecting regions free of detectable galaxy emission. These regions were selected using circular apertures with radii of 10–15 pixels, avoiding contamination from nearby sources. For each cube, we computed a median background spectrum, which was then fitted with a high-order polynomial (degree 9–13) to capture large-scale spectral variations. The resulting model was subtracted from the data cube.

To ensure a consistent spatial resolution across wavelengths, we homogenized the point-spread function (PSF) of the NIRSpec data, convolving to a common spatial resolution corresponding to the PSF at observed-frame $\sim 5~\mu$m. The PSF was characterized using a star observed within the GA-NIFS program (see Appendix~\ref{app:A}). We modelled it with an elliptical 2D Gaussian profile, allowing for wavelength-dependent variations of the semi minor and major axes 
while keeping the position angle as a free parameter (\citealt{DEugenio2024, Jones2026}). For our targets, the spatial sampling corresponds to $\sim 200-300$~pc, and the spatial resolution to $\sigma \sim 400-600$~pc.

To account for the wavelength-dependent aperture losses we rescale the R100 spectrum to the NIRCam photometry by fitting a polynomial correction factor within our SED modelling, using a 0.5\arcsec\ radius aperture (see Appendix~\ref{app:B}).

\subsection{Apertures for spatially integrated spectra}\label{sec:apertureforintegratedspectra}

For each galaxy, we constructed a global spatial aperture used both for extracting integrated spectra and for defining the region for the spatially resolved analysis (see left panel in Figs.~\ref{fig:10578-GS}-\ref{fig:12306-COS}). The apertures are designed to retain only regions where the stellar continuum is robustly detected.

We first estimated the continuum at different wavelengths in the rest-frame reference system. We computed it as the mean in narrow windows (0.02~$\mu$m) centred at specific rest-frame wavelengths (at 0.25, 0.55, and 0.75 $\mu$m) corresponding to regions free of emission lines. Due to its higher redshift, for EGSY8p7 we only adopted the band at 0.25~$\mu$m.
The corresponding noise maps were computed as the standard deviation within the same narrow band, allowing us to derive the S/N maps of the continuum for each band. The final apertures were then defined by selecting spaxels for which the S/N exceeds 3 in at least one of the continuum bands considered. 
The uncertainties on the integrated spectra were propagated by summing in quadrature the errors in each spaxel. However, since the spaxels are not independent due to the PSF, this approach underestimates the true uncertainties. We therefore rescale the errors to match the standard deviation in the continuum.

\begin{figure*}[th]
    \centering
    \includegraphics[width=1\linewidth]{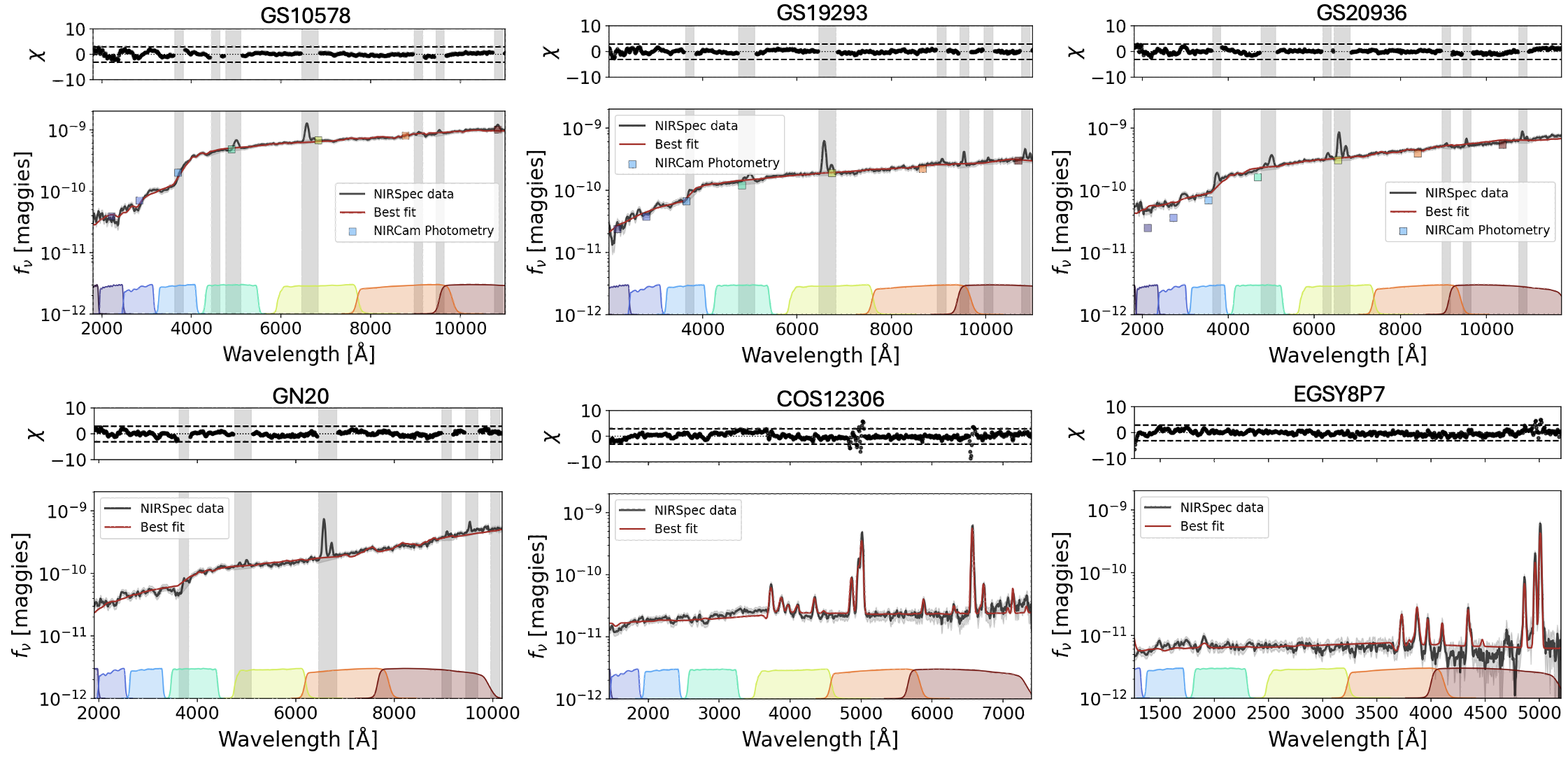}
    \caption{Integrated spectra extracted from the apertures showed in the left panels of figures in Appendix \ref{app:C}. Data are in black, best fit with non-parametric SFH model in red, and grey bands represent the masked emission lines. Photometric points refer to a circular aperture of radius 0.5" (CIRC6, JADES Data Release 5, \citealt{Johnson2026, Robertson2026}). The photometric points for GS20936 lie below the spectrum because the spectrum was extracted from an aperture $\sim$2 times larger than the photometric aperture (see Fig.\ref{fig:20936-GS}). The upper panels show the residuals (defined as (data--model)/error), with horizontal dashed black line showing $\chi=\pm3$  }
    \label{fig:unresolved_spec}
\end{figure*}

\subsection{SED fitting}\label{sec:Analysis_Results}
To model the stellar continuum, we adopt the \texttt{Prospector} SED-fitting framework \citep{Johnson2021}. The code combines stellar population synthesis models from \texttt{fsps} \citep{Conroy2009,Conroy2010} with both parametric and non-parametric descriptions of the SFH, while accounting for additional physical processes such as dust attenuation, nebular emission, and the presence of an AGN. We configured \texttt{fsps} to use the MILES stellar atmospheres (\citealt{Falcon-Barroso2011}) and MIST isochrones (\citealt{Choi2016}). We constructed three families of SFH models:

\begin{itemize}
    \item \textbf{Parametric SFH:} In the parametric models, we adopted a delayed-$\tau$ SFH, in which the SFR evolves as
%
$\mathrm{SFR}(t) \propto t \, \exp(-t/\tau)$,
where $t$ is the time since the onset of star formation ($t_{\rm age}$) and varies between 0 and $t_{\rm age}$, and $\tau$ is the characteristic star-formation timescale. We allow $t_{\rm age}$ to vary between the time of observation and $z=15$ \citep[for a similar approach see][]{Marconcini2024,Marconcini2025}.

    \item \textbf{Parametric SFH + burst:} In this model, we introduced an additional burst of star formation on top of the parametric SFH. 
    The time of the burst (\textit{tburst}) is free to vary between the time of observation and $z = 15$, while the fraction of stellar mass formed during the burst (\textit{fburst}) is allowed to vary between 0 and 80\% of the total formed stellar mass (e.g. \citealt{Suess2022}).

    \item \textbf{Non-parametric \texttt{continuity\_sfh}:} 
    In this case, the SFH is described by 9 time bins between the redshift of the galaxy and $z = 15$: the first three bins correspond to $0-10$~Myr, $10-30$~Myr, and $30-100$~Myr, to capture variation in the recent SFH, while the remaining six bins are logarithmically spaced in time (e.g. \citealt{Marconcini2024}). 
    We adopt a uniform prior on the SFR ratios between adjacent time bins ($-2 < \log (\mathrm{SFR}_{i}/\mathrm{SFR}_{i+1}) < 2$).
    The advantage of this non-parametric approach is that it does not impose a predetermined functional form for the SFH and therefore allows the reconstruction of complex evolutionary histories, including rejuvenation events and quenching phases. The recent star formation activity is summarized by the average SFR over the last 10 and 100 Myr (SFR$_{10}$ and SFR$_{100}$). 

\end{itemize}
Simple stellar population spectra were generated consistently with the adopted SFH in each model, assuming an initial mass function (IMF) from \cite{Kroupa2001} with lower and upper mass cutoffs of $0.08\,M_\odot$ and $120\,M_\odot$, respectively. A 5\% error floor is added to the flux uncertainties to account for the uncertainties in the underlying stellar models following \cite{Nersesian2025}. For simplicity, we modeled the dust attenuation using a power law  with a slope of $-0.7$, following \citet{Charlot2000}, applied uniformly to both young and old stellar populations. In other words, we adopt a single-component attenuation model by setting the \texttt{dust1} parameter in \texttt{Prospector} to zero.

For GS10578, GS20936, GS19293, and GN20, the emission lines may include a significant contribution from the AGN. 
Given the difficulty of decoupling the AGN and star formation contributions to the emission line, we therefore masked the emission lines during the fitting procedure for these four sources.
For the starburst galaxies COS12306 and EGSY8p7, the fit was performed including nebular emission lines in the model. In this case, the nebular emission was generated using a \texttt{CLOUDY} \citep{Ferland1998,Ferland2013} grid within \texttt{FSPS}, following the prescription of \citet{Byler2017}. 
We allowed the ionization parameter to vary within
$-4.0 < \log U < -1.0$,
while the gas-phase metallicity was tied to the stellar metallicity. We assumed the same dust attenuation 
for both the stellar and nebular components.

Finally, we did not include an AGN component in the modelling, as the sources are either starburst or Type 2 AGN for which the continuum emission is expected to be obscured. In GN20 there may be some contribution but negligible in the optical band (Hamed in prep.). We also neglected dust emission, since the spectra extend only up to 1.2 $\mu$m, where the AGN torus contributes only marginally (\citealt{DEugenio2024}).

A summary of the explored parameter space in the SED fitting is given in Table~\ref{tab:setup}. 
In the unresolved fit, we sampled the posterior distributions with Dynesty (\citealt{Speagle2020}; see details in the Table \ref{tab:setup}), while for the spatially resolved analysis we used a Levenberg-Marquardt optimization, useful for non-linear problem for a fast convergence while maintaining a low mean square errors compared to other optimization algorithms. 


\section{Results and discussion}\label{sec:Results}

In this section, we present the results of both the unresolved and spatially resolved analyses. We begin in Sect.~\ref{sec:unresolved_fits} by comparing the results obtained with different models applied to the spatially integrated spectra, and by assessing how the inferred physical properties vary under different assumptions on the SFH. 
Section~\ref{sec:integratedquantities} compares integrated physical quantities (M$_\star$, SFH) derived from spatially integrated spectra with those obtained by summing the spatially resolved measurements over the same apertures. 
Section~\ref{sec:radialtrends} focuses on radial variations in the stellar population properties. We first investigate spatial trends in the SFH and the build-up of stellar mass, to assess evidence for inside-out growth (Sect.~\ref{sec:sfh_radius}). We then compare the evolution of the sSFR in nuclear and off-nuclear regions to explore differences in relative growth and determine whether distinct parts of the galaxies evolve along the main sequence or experience phases of suppressed or enhanced star formation (Sect.~\ref{sec:nucleus_vs_outer}).

In Sect.~\ref{subsec:outflows}, we discuss the possible physical drivers of the observed suppression of star formation, with particular emphasis on the role of the outflow. Finally, Sect.~\ref{sec:rejuvenation} presents the case of GS20936, which shows evidence for galaxy rejuvenation after a phase of reduced star formation activity. 

\begin{figure}
    \centering
    \includegraphics[width=0.98\linewidth, trim=0mm 8mm 0mm 0mm,clip]{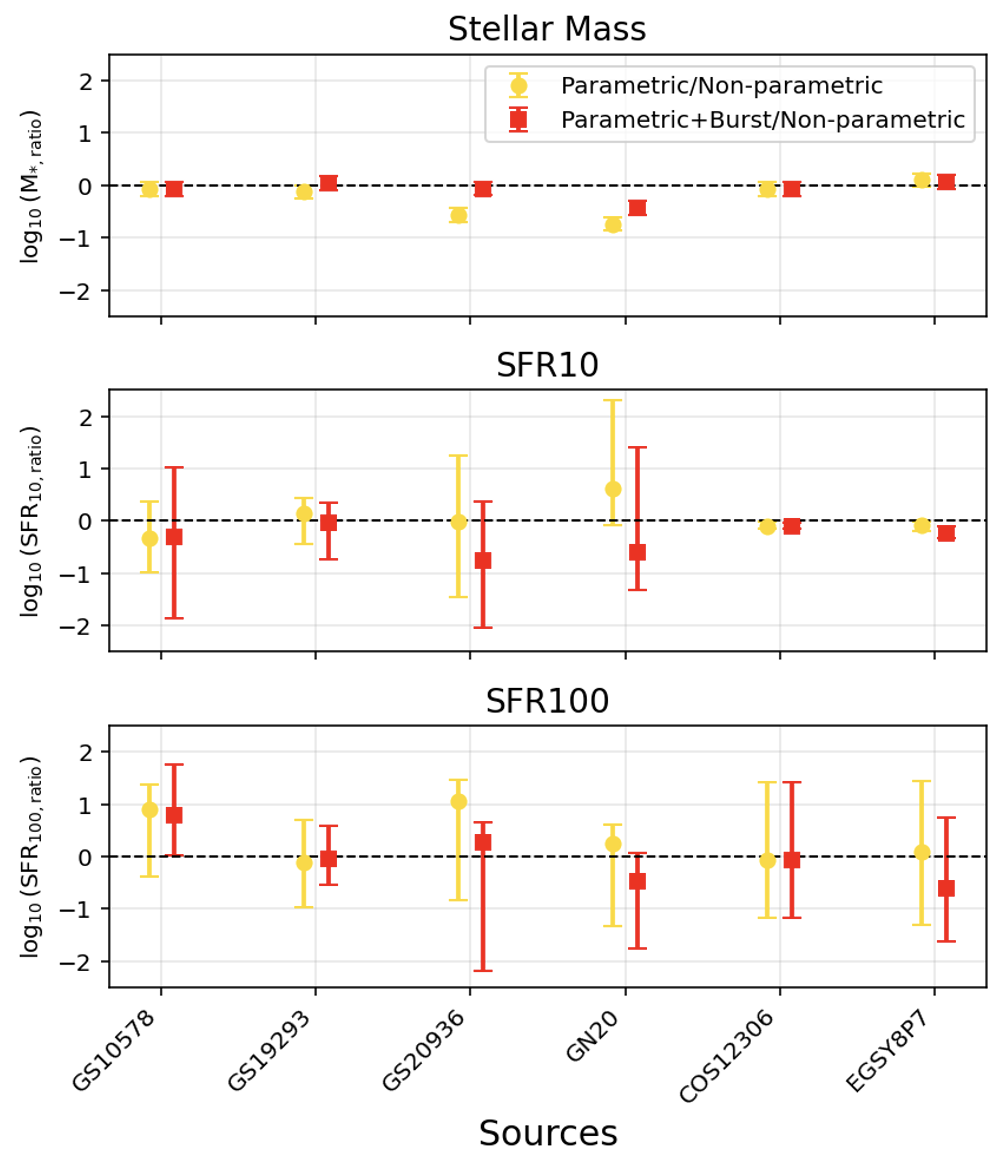}
    \caption{Ratios of the stellar mass (top), SFR10 (middle) and SFR100 (bottom) derived from different fitting methods, as labeled in the top panel. 
    Error bars indicate the propagated uncertainties. }
    \label{fig:ratio}
\end{figure}

\begin{figure*}
    \centering
    \includegraphics[width=1\linewidth]{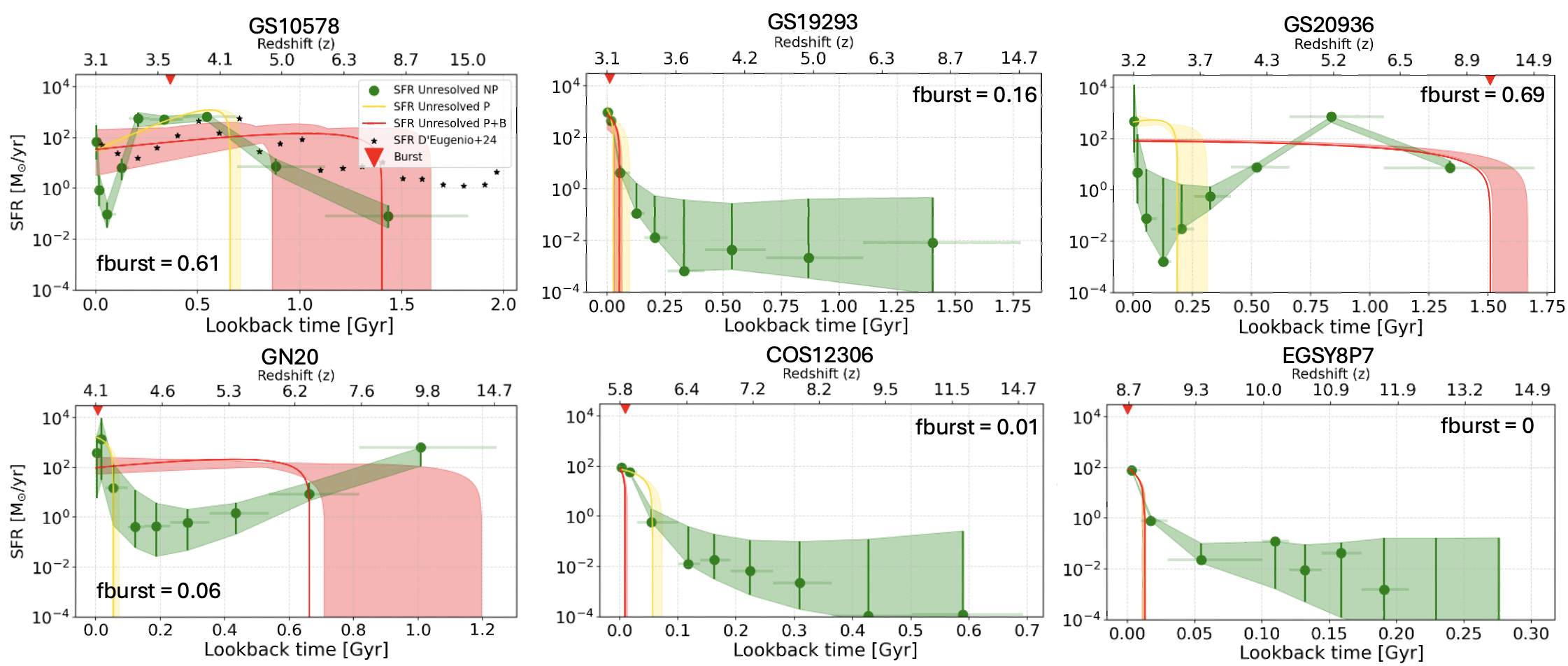}
    \caption{SFHs derived from integrated spectra using three different methods: non parametric (NP, green), delayed parametric (P, yellow) and bursty parametric method (P+B, red). The red triangles indicate the epoch of the burst. The black stars in GS10578 panel show the unresolved SFH derived in \cite{DEugenio2024}. Shaded areas and vertical errorbars represent 16th and 84th percentiles intervals; horizontal bars identify the temporal bins. fburst denotes the fraction of the total stellar mass formed during the burst. }
    \label{fig:sfh_unresolved}
\end{figure*}

\begin{figure*}[h!]
    \centering
    \includegraphics[width=1\linewidth]{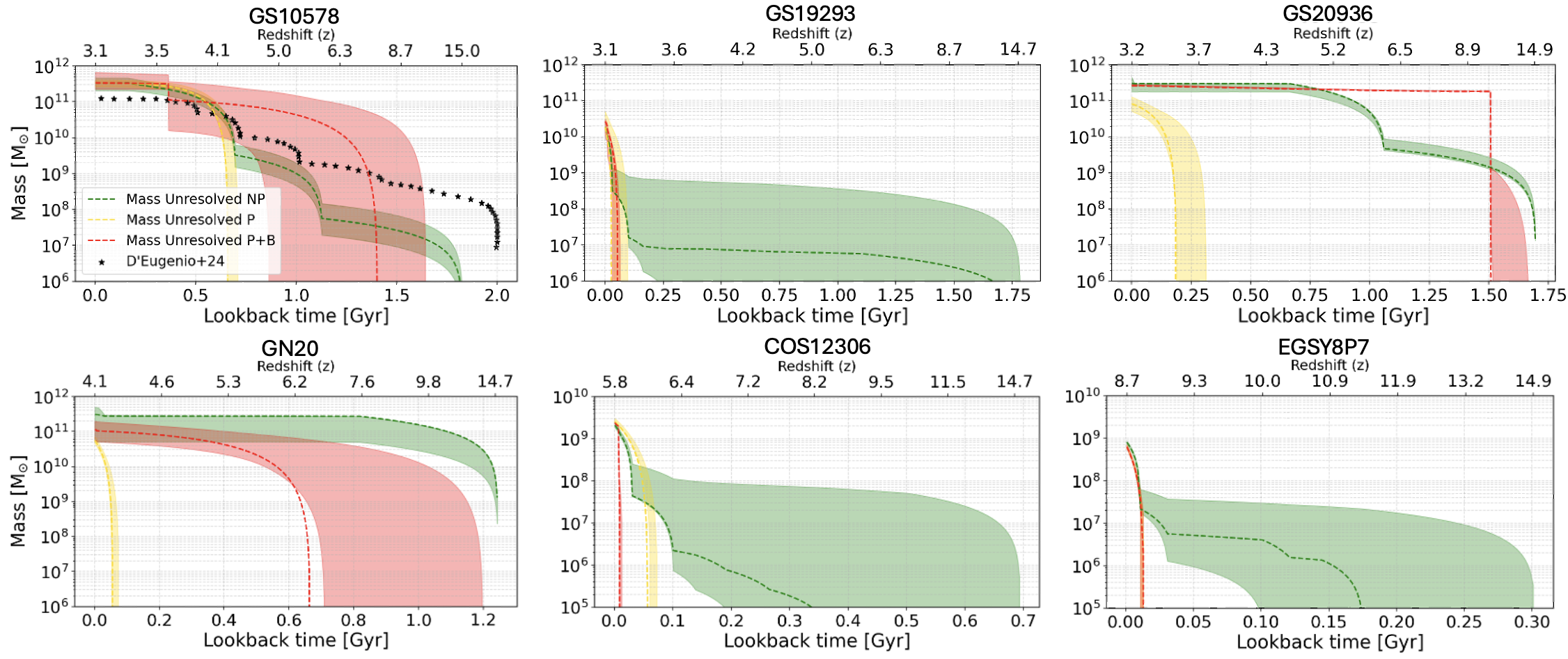}
    \caption{Cumulative stellar masses derived from integrated spectra using three different methods, as in Fig.~\ref{fig:sfh_unresolved}.}
    \label{fig:mass_unresolved}
\end{figure*}

\begin{table*}
\caption{Stellar masses and SFRs averaged over the last 10 and 100 Myr, derived from integrated spectra using different SFH parameterisations. }
\label{tab:integrated_results}
\centering
\resizebox{\textwidth}{!}{
\begin{tabular}{lcccc|cccc|cccc}
\hline
& \multicolumn{4}{c}{Parametric SFH}
& \multicolumn{4}{c}{Parametric + Burst SFH}
& \multicolumn{4}{c}{Non-parametric SFH} \\
\hline
Target &
$M_\star$ &
SFR$_{10}$ &
SFR$_{100}$ &
$\chi^2_{\mathrm{red}}$ & 
$M_\star$ &
SFR$_{10}$ &
SFR$_{100}$ &
$\chi^2_{\mathrm{red}}$ & 
$M_\star$ &
SFR$_{10}$ &
SFR$_{100}$ &
$\chi^2_{\mathrm{red}}$ \\
&
$[10^{10}\,M_\odot)$] &
$[M_\odot\,\mathrm{yr}^{-1})$] &
$[M_\odot\,\mathrm{yr}^{-1})$] &
&
$[10^{10}\,M_\odot)$] &
$[M_\odot\,\mathrm{yr}^{-1})$] &
$[M_\odot\,\mathrm{yr}^{-1})$] &
&
$[10^{10}\,M_\odot)$] &
$[M_\odot\,\mathrm{yr}^{-1})$] &
$[M_\odot\,\mathrm{yr}^{-1})$] &
\\
\hline

GS10578 &
$33 \pm 5$ & $32 \pm 5$ & $48 \pm 7$ & 0.58 &
$33 \pm 5$ & $34 \pm 6$ & $37 \pm 7$ & 0.57 &
$39 \pm 6$ & $69 \pm 10$ & $6 \pm 1$ & 0.56 \\

GS19293 &
$1.8 \pm 0.3$ & $1100 \pm 200$ & $190 \pm 30$ & 0.8 &
$2.6 \pm 2$ & $736 \pm 90$ & $225 \pm 80$ & 0.78 &
$2.4 \pm 0.4$ & $810 \pm 120$ & $250 \pm 40$ & 0.77\\

GS20936 &
$8.2 \pm 1.2$ & $450 \pm 70$ & $510 \pm 80$ & 0.9 &
$26 \pm 4$ & $83 \pm 12$ & $82 \pm 12$ & 0.65 &
$30 \pm 5$ & $470 \pm 70$ & $44 \pm 7$ & 0.56 \\

GN20 &
$5.5 \pm 0.8$ & $1600 \pm 240$ & $560 \pm 80$ & 0.88 &
$11 \pm 1.3$ & $97 \pm 10$ & $110 \pm 15$ & 0.61 &
$30 \pm 5$ & $390 \pm 60$ & $320 \pm 50$ & 0.71 \\

COS12306 &
$0.25 \pm 0.04$ & $220 \pm 30$ & $77 \pm 12$ & 1.59 &
$0.2 \pm 0.5$ & $51 \pm 9$ & $4 \pm 1$ & 1.4&
$0.20 \pm 0.03$ & $87 \pm 13$ & $20 \pm 3$ & 1.57 \\

EGSY8p7 &
$0.62 \pm 0.09$ & $120 \pm 20$ & $12 \pm 2$ & 0.82 &
$0.61 \pm 0.08$ & $61 \pm 10$ & $6 \pm 2$ & 0.82 &
$0.72 \pm 0.11$ & $77 \pm 12$ & $7 \pm 1$ &0.96 \\

\hline
\end{tabular}
}
\end{table*}


\subsection{Comparison of global M$_*$ and SFHs derived from parametric and non-parametric approaches.}\label{sec:unresolved_fits}

Figure~\ref{fig:unresolved_spec} presents the integrated \prism spectra extracted within the defined apertures  (Figs.~\ref{fig:10578-GS}-\ref{fig:12306-COS}). 
In GS10578, we observe a pronounced Balmer break, indicative of a stellar population dominated by A-type stars with typical ages of $\sim0.1-1$~Gyr (\citealt{DEugenio2024}). This feature is also present, albeit less prominent, in the other galaxies exhibiting AGN signatures (GN20, GS19293 and GS20936). In contrast, the spectra of the two starburst galaxies (COS12306 and EGSY8p7) do not show Balmer breaks and are characterized by a much weaker continuum, with the emission lines dominating the SED.

We quantified the overall quality of the SED modeling by computing the reduced $\chi^2_{\mathrm{red}}$.
We found values of $\chi^2_\mathrm{red}$ in the range 0.56 -- 1.6, indicating a high-quality overall fit for all the used models, with general better values for the non-parametric model. For this reason, in Fig.~\ref{fig:unresolved_spec} we presented the fit obtained using the non-parametric approach (red curve).
The stellar masses, recent SFRs, together with the $\chi^2_{\mathrm{red}}$ inferred from these fits are summarized in Table~\ref{tab:integrated_results}. 

To better compare the results of the three different models, we show in
Fig.~\ref{fig:ratio} the ratios of stellar mass, SFR$_{10}$, and SFR$_{100}$ derived from parametric and non-parametric SFH models (gold), as well as the ratios of the same quantities derived with the parametric models including bursts and the non-parametric model (red).
The non-parametric models generally yield higher stellar masses than the parametric SFH models, while showing better agreement with the parametric+burst solutions.  
The inferred SFR$_{10}$ and SFR$_{100}$ values are broadly consistent (<3$\sigma$) across the different model assumptions, although for some targets (GN20 and COS12306) the parametric models tend to underestimate SFR$_{100}$ relative to the non-parametric results. 
We note that for the sources in which we did not fit the emission lines, as probably contaminated by the AGN, the uncertainties on the SFR$_{10}$ and SFR$_{100}$ ratios are quite large (up to 1dex).

Figures~\ref{fig:sfh_unresolved} and \ref{fig:mass_unresolved} present the global SFHs and stellar mass assembly histories of our targets, respectively, derived using the three different modelling approaches. 
For GS10578, GS20936, and GN20, the SFH exhibit episodes of both increasing and decreasing SFR, whereas for GS19293 and the two z>5 starburst galaxies (EGSY8p7 and COS12306), the SFH is monotonic, increasing over time.
In the following, we briefly discuss the main results for each galaxy.

{\noindent \normalsize\sffamily GS10578.}
The non-parametric SFH shows an early rise in star formation, peaking at $z \sim 3.3-4$ with SFR $\sim 1000$~M$_{\odot}~\mathrm{yr^{-1}}$, followed by a phase with lower activity lasting until $\sim 10~\mathrm{Myr}$ ago, when star formation appears to increase again. The parametric+burst model reproduces this behaviour, with the burst occurring at a similar epoch of the peak of SFR in the non-parametric SFH, and contributing for around 61\% of the total mass. In contrast, the purely parametric model captures the recent evolution but fails to account for older stellar populations. The SFH of GS10578 was previously studied by \citet{DEugenio2024} considering the unresolved multi-wavelength UV-to-NIR photometry and using CIGALE (\citealt{Boquien2019}). We find an overall consistent SFH and we infer a stellar mass approximately a factor of two higher, likely due to differences in the adopted modelling framework. All the approaches suggest  that the galaxy has built up most of its mass  by $z=3.5$, that is before the onset of the  episode of SF suppression.

{\noindent \normalsize\sffamily GS19293.}
Both parametric and non-parametric models provide a stellar mass of $\sim 2\times10^{10}M_{\odot}$ (consistent with \citealt{Venturi2025}), and indicate that GS19293 is a young system, with significant SF occurring within the last $\sim 250~\mathrm{Myr}$. All models point to a recent (<100~Myr) phase of intense star formation, with SFR $\sim 10^{2}-10^{3}M_{\odot}~\mathrm{yr^{-1}}$. However, the non-parametric model shows an excess in the SFR at early epochs compared to the parametric case, suggesting earlier formation of the system. Including a burst component in the parametric SFH does not significantly alter the inferred SFH, as the burst contributes only 16\% of the total mass.

{\noindent \normalsize\sffamily GS20936.}
The inferred SFH is qualitatively similar to that of GS10578, with an early phase of star formation peaking at $z \sim 5$, followed by a period with lower activity. It resumes in the last $\sim 150 ~\mathrm{Myr}$, reaching SFRs above $400~M_{\odot}\,\mathrm{yr^{-1}}$ in the most recent $\sim 10~ \mathrm{Myr}$. This suggests a rejuvenation phase after a period of suppressed activity. Before the onset of this phase of reduced SF ($z\simeq5$),  the galaxy has already  built  up most of its mass, as was the case for GS-10578. The parametric+burst model suggests an early burst that accounts for $\sim 69$\% of the total stellar mass. 

{\noindent \normalsize\sffamily GN20.}
The inferred stellar mass spans a wide range $\sim (5-30)\times10^{10}$~M$_{\odot}$, depending on the adopted SFH model. The galaxy exhibits high recent SFRs, up to 1600~M$_{\odot}\,\mathrm{yr^{-1}}$. These measurements are consistent with previous results from \citet{Tan2014} and \citet{Colina2023}, who derived M$_\star = 10^{11}$~M$_{\odot}$ and $5\times10^{10}$~M$_{\odot}$, and SFR~$= 1860$ and 524~M$_{\odot}\,\mathrm{yr^{-1}}$, respectively, from integrated SED analyses. The non-parametric model suggests an early formation phase followed by a period of reduced activity lasting $\sim 500~\mathrm{Myr}$, and a subsequent recent starburst. In contrast, the purely parametric model favors a much younger system, highlighting the strong dependence of inferred evolutionary histories on the adopted SFH parameterization. 

{\noindent \normalsize\sffamily COS12306.}
The inferred stellar masses are consistent across models ($\sim$ 2--4 $\times10^{9}$~M$_{\odot}$). 
All the models indicate a recent formation scenario, with a consistent SFR ($\sim100~$~\Msunyr) over the last 10~Myr. The contribution from the burst is only 1\% of the total mass, while also in this case the non-parametric model accounts for a non-zero SF at an earlier epoch. 

{\noindent \normalsize\sffamily EGSY8p7.}
The estimates of stellar mass ($\sim 7\times10^{9}M_{\odot}$) are consistent across models and agree with previous results \citep{Larson2023}. All SFH prescriptions indicate a very young system, with star formation beginning less than $\sim 20-50$ Myr ago, even if the non-parametric model allows for a modest level of star formation at an earlier epochs, as in GS19293. The recent SFR is robust across the different models, ranging between $\sim 95$ and $150~M_{\odot}\,\mathrm{yr^{-1}}$, in agreement with \citet{Marques-Chaves2024}. No evidence for quenching is found, consistent with the extremely young age of the galaxy. Differences between H$\beta$- and UV-based SFRs reported in the literature \citep{Zamora2025} further suggest a rapidly varying star formation activity, as H$\beta$ traces star formation on shorter timescales ($\lesssim 10$ Myr), whereas the rest-frame UV is sensitive to star formation averaged over longer periods ($\sim 100$ Myr).



Detailed SFHs can vary significantly across different modelling approaches due to their underlying assumptions and constraints, with the largest discrepancies typically arising at early epochs (e.g. \citealt{Leja2019, Lower2020, Tacchella2022, Suess2022, Whitler2023, GimenezArteaga2024, Wang2025}). 
In particular, we showed above that parametric SFHs are too rigid and tend to underestimate the role of old stellar population, yielding to a lower mass estimation as reported in several works (\citealt{Ciesla2017, Iyer2017, Lower2020}). 
For these reasons, we chose to perform our resolved analysis using a non-parametric SFH.


\subsection{Integrated versus resolved global stellar measurements}\label{sec:integratedquantities}

To assess the impact of spatial resolution on the inferred global stellar population properties, we compare measurements derived from spatially integrated spectra with those obtained by summing the results of the spaxel-by-spaxel analysis over the same apertures. The latter is presented in Appendix~\ref{app:C}. This comparison allows us to quantify systematic biases introduced by unresolved observations, particularly in systems hosting spatially distinct stellar populations with different ages and dust attenuation. We first focus on stellar mass estimates and evaluate the role of the outshining effect in the context of previous observational studies (Sect.~\ref{sec:outshining}). We then compare the SFHs inferred from resolved and unresolved analyses, highlighting how spatial information affects the reconstructed assembly histories of the galaxies (Sect.~\ref{sec:sfh_comparison}).

\subsubsection{The outshining effect: impact on stellar mass estimates}\label{sec:outshining}
Comparing the mass obtained by summing the spaxel-by-spaxel measurements with that derived from the integrated spectrum reveals a discrepancy, with the former generally yielding a larger value.
This is a known effect, and is defined in literature as ‘outshining’ (\citealt{Trager2008,Jain2024,Narayanan2024,Hamed2026}). Outshining is thought to occur as a consequence of the fact that young and bright stars such as O and B-type stars in integrated spectra can be dominant compared to the older (>100 Myr) stellar populations. The low-mass stars in fact dominate the mass budget but contribute little to the observed rest-UV/optical light (\citealt{Schechter1976,Chabrier2000,Kroupa2002}) or in other words have a much higher $M_\star/L$ than young stars. This leads to an underestimation of the stellar mass in a galaxy when derived from integrated spectra. 

\begin{figure}
    \centering
    \includegraphics[width=1\linewidth]{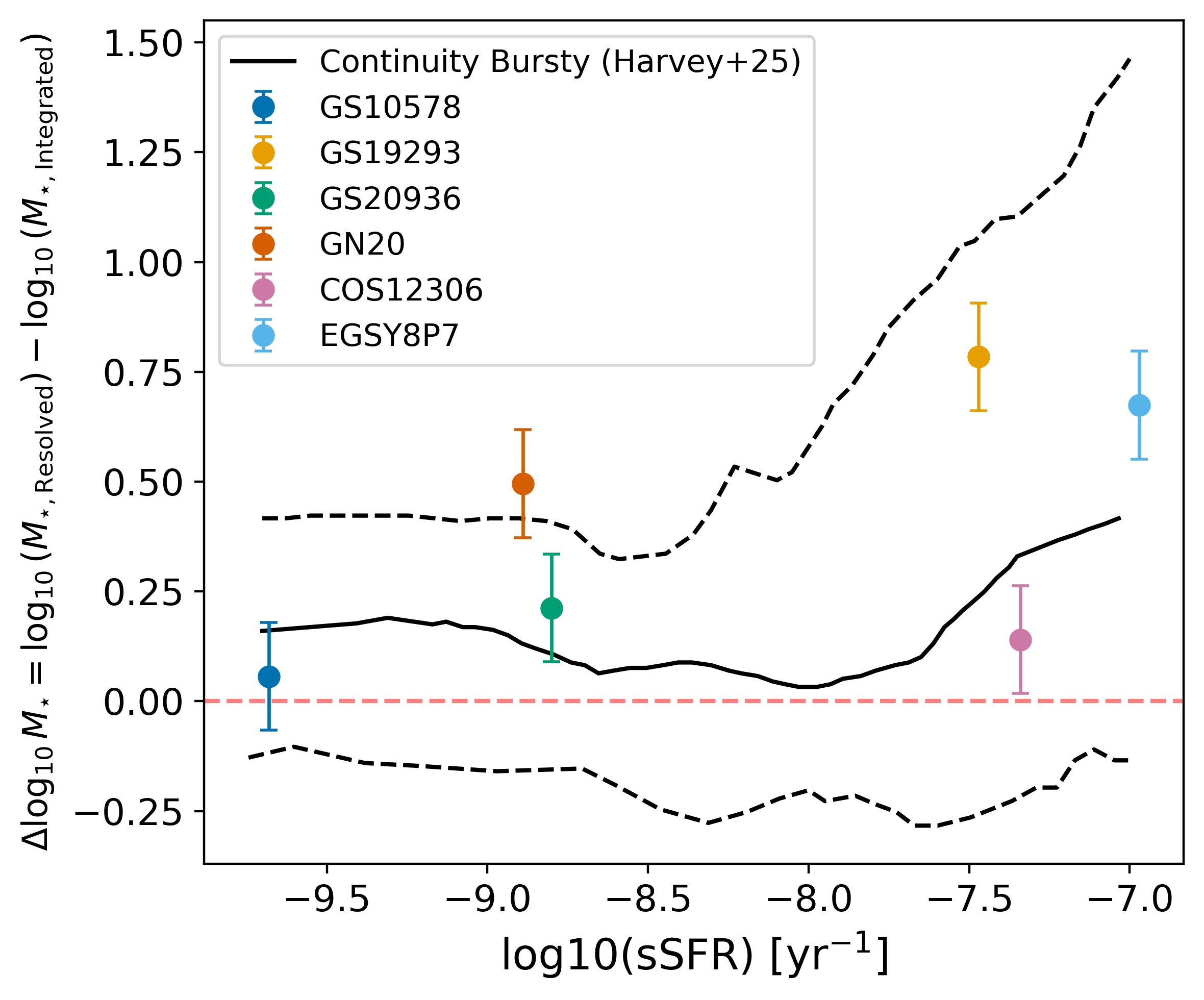}
    \caption{Stellar mass offset ($\Delta M_\star = \log_{10} M_{\star,\rm Resolved} - \log_{10} M_{\star,\rm Integrated}$) as a function of sSFR. The latter is computed as the ratio between the SFR in the last 10 Myr and the stellar mass derived from the unresolved spectrum. The black curve represents the relation (and associated scatter in dashed) derived by \citealt{Harvey2025} (their Fig.~7).  }
    \label{fig:outshining}
\end{figure}

To quantify this effect, we plot in Fig. \ref{fig:outshining} the stellar mass offset defined as $\Delta M_\star = \log_{10} (M_{\star,\rm Resolved}) - \log_{10} (M_{\star,\rm Integrated})$ as a function of sSFR, the latter calculated as the SFR in the last 10 Myr divided by the stellar mass from the unresolved analysis. 
The stellar mass offset spans a broad range, from 0.1~dex up to around 0.75~dex. 
\cite{Harvey2025} show that the mass offset increases with sSFR, although with large scatter, with galaxies undergoing recently rising SFHs exhibiting systematically larger offsets. 
Given the limited size of our sample, the statistics remain poor and the relation exhibits significant scatter, although consistent with the dispersion reported by \citealt{Harvey2025}. The only exception is GN20 that might be out of the relation due to the high obscuration and an overestimation of the resolved stellar mass.  
Interestingly, GS10578 exhibits the lowest stellar mass offset in our sample, consistent with its quiescent nature and its low cold molecular gas content (\citealt{Scholtz2026}). GS19293, on the other hand, shows a large stellar mass offset, probably due to its high recent burst of star formation dominating the blue integrated spectrum and leading to an underestimation of the stellar mass.
This result highlights the importance of spatially resolved data in mitigating systematic biases (see also \cite{Sorba2018, GimenezArteaga2024,Narayanan2024,Hamed2026} ). 

\begin{figure*}
    \centering
    \includegraphics[width=1\linewidth]{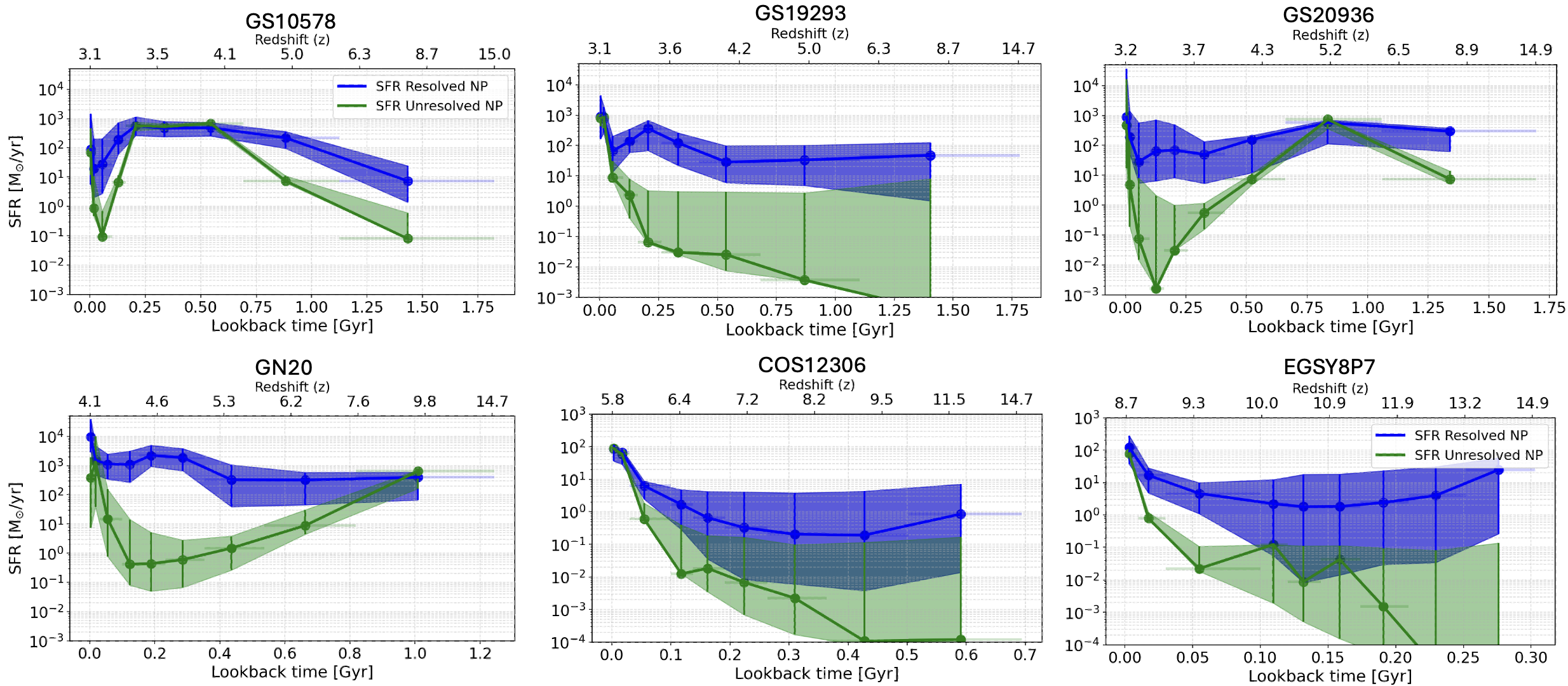}
    \caption{Spatially resolved and unresolved SFH derived using the non-parametric (NP) approach. SFH derived from integrated spectra (green) and summing the SFR contribution of each pixel (blue curve).  }
    \label{fig:resolved_unresolved}
\end{figure*}

\begin{figure*}
    \centering
    \includegraphics[width=1\linewidth]{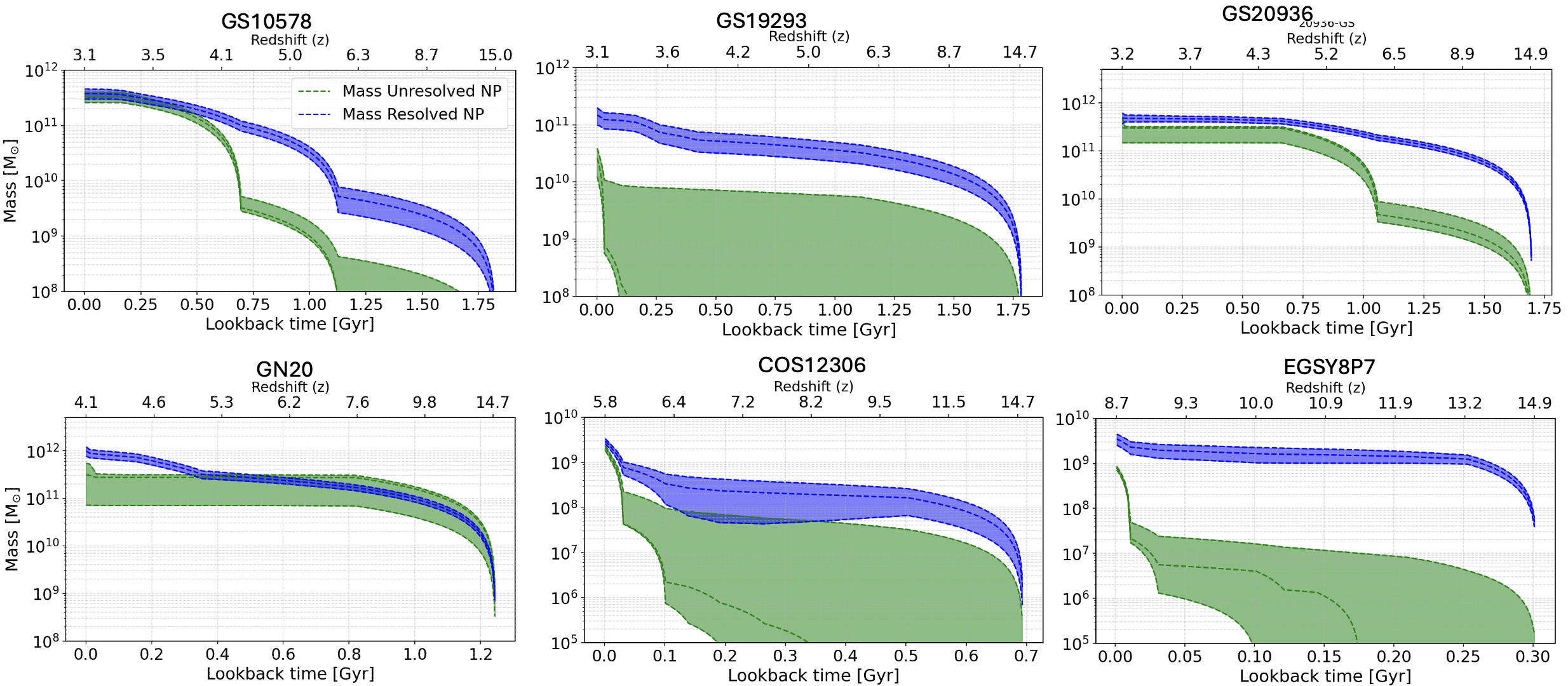}
    \caption{Spatially resolved (blue) and unresolved (green) cumulative stellar mass assembly history obtained with the non-parametric SFH method for the targets in our sample.  }
    \label{fig:resolved_unresolved_mass}
\end{figure*}

\subsubsection{Comparing resolved and unresolved SFHs}\label{sec:sfh_comparison}

Figure~\ref{fig:resolved_unresolved} compares, for each target, the SFH obtained by adding the SFR in each time bin for all spatial pixels with that derived from the unresolved spectrum analysis; similarly, Fig.~\ref{fig:resolved_unresolved_mass} shows the comparison for the global stellar mass assembly history. For all galaxies the SFR from the resolved method are higher (or equal) than the ones from the unresolved analysis.  This is also reflected in the higher stellar mass assembly histories for the resolved method. For most objects, the SFH inferred from integrated spectra and that derived from resolved analyses differ mostly in early epochs, where it is more difficult to constrain the SFR. In contrast, the two approaches show good agreement over the last few million years, where the SFR can be estimated more reliably.  Similar behaviour is observed comparing the global stellar mass assembly histories.

For GS10578, the two SFHs show similar behaviour. The SFR increases from the onset of star formation up to a $z = 3-4$, and subsequently declines over a period of $\sim $200 Myr. In the last few Myrs, there appears to be a small increase in star formation. 
The resolved analysis indicates that GS10578 had already assembled 
a stellar mass of $\sim10^{10}$~M$_\odot~$ at $z \sim 6$, whereas the unresolved reconstruction reaches the same stellar mass only by $z\sim 4.2$.

In GS19293, the SFHs show constant SFRs at $z>4$, although with a substantial difference between resolved and unresolved cases: in the latter, the SFR remains $< 5$~\Msunyr at $z\gtrsim 3.5$, while in the resolved case, the SFR is $> 10$~\Msunyr since early epochs. 
From $z\sim 4$ down to the redshift of the galaxy, the SFR rises in both cases, but more steeply in the unresolved analysis. 
These difference in SFHs are reflected in the stellar mass assembly: in the resolved case, most stars form at early times, while in the unresolved case the mass is assembled predominantly in the last few 10s Myr.

In GS20936, the SFHs follow a similar overall trend, with SFRs $>10$~\Msunyr already in place at $z> 5.2$. In both cases, the SFR decreases over $\sim$ 700 Myr (from $z = 5.2$ to $\sim 3.4$), with stronger suppression in  the unresolved SFH. The mass already exceeds $5\times10^{10} M_\odot~$ at $z\sim5.5$ in both cases. 


\begin{figure*}
    \centering
    \includegraphics[width=1\linewidth]{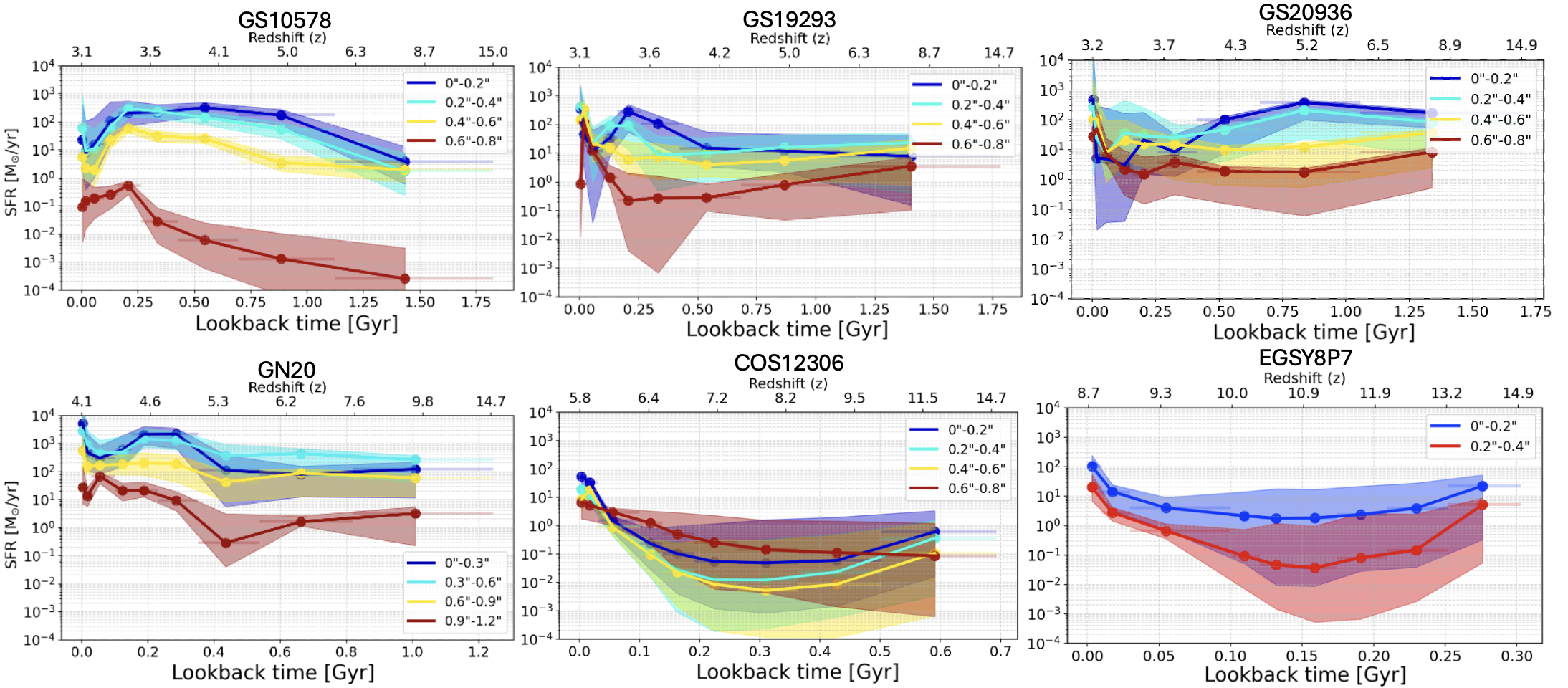}
    \caption{SFR as a function of the lookback time. The colour coding indicates the distance from the galaxy centre, as labeled in each panel. }
    \label{fig:sfh_radius}
\end{figure*}

The most pronounced discrepancy between resolved and unresolved SFHs is observed in GN20, as expected given its complex and well-resolved structure. In both cases, star formation begins early, with SFRs $>100$~\Msunyr at $z\sim 10$. From this point, the two diverge: the unresolved SFH shows low SFR levels ($< 10$~\Msunyr) lasting $\sim 750$~Myr, while the resolved SFH continues to rise. This difference likely reflects the outshining effect. GN20 hosts at least two distinct components: a heavily obscured nucleus (A$_V\sim 7$) and a bright, arc-shaped star-forming region with low extinction (Hamed et al., in prep.), the latter dominating the unresolved spectrum.

In the remaining two galaxies (COS12306 and EGSY8p7), the trends of SFH are very similar, with a smoother (but also more uncertain) increase in SFR compared to the other sources in our sample. The SFHs reach a peak of $\sim 100~$\Msunyr at the reshift of the galaxies. 
While the unresolved SFHs are dominated by very recent bursts, with little contribution from stars older than $\sim 50$~Myr, the resolved analysis indicates a more continuous star formation over extended timescales. 
This discrepancy is consistent with the outshining effect (see e.g. Fig. 10 of \citealt{GimenezeArteaga2023} for similar results).

Summarising, the most massive galaxies hosting AGN (GS10578, GS20936, GS19293, GN20) exhibit more variable SFR trends with phases of diminished star formation, distinct from those of lower-mass, higher-$z$ starburst systems COS12306 and EGSY8p7, whose evolution appears more steadily rising and lacks clear signs of sudden decreases or interruptions of star formation. 
This suggests that AGN activity may play an important role in shaping the star formation histories of massive galaxies at early cosmic epochs. We further investigate this scenario in the following section.



\subsection{Radial trends in SFHs}\label{sec:radialtrends}

The spatially resolved analysis allows us to investigate how star formation proceeded across different regions of each galaxy. In this section, we examined radial variations in the SFH to assess whether stellar mass assembly follows an inside-out pattern. We then compared the evolution of the sSFR in nuclear and off-nuclear regions to explore differences in the relative growth of central and outer components.

\subsubsection{Spatially resolved SFH: Evidence for inside-out growth}\label{sec:sfh_radius}
Radial profiles offer insight into the buildup of stellar mass and the spatial distribution of star formation across different cosmic epochs (\citealt{Belfiore2018,Haryana2025}).
This analysis is also important for understanding the possible causes of the discrepancy between the SFH derived from the unresolved spectrum and the  SFH derived from the spatially resolved analysis.  

We constructed concentric circular annuli centred on the galaxy nucleus (defined as the position of the continuum-emission peak at 5100\AA, see black stars in Figs. \ref{fig:10578-GS}-\ref{fig:12306-COS}), with bin widths of $\sim 2\sigma_{\rm PSF}$ (i.e. $\delta r = 0.2''$, $0.9\mathrm{kpc}<r<2\mathrm{kpc}$ in the probed redshift range), except for GN20, where we considered 3$\sigma_{\rm PSF}$ annuli (i.e. $\delta r = 0.3''$) due to its larger spatial extent. As a reference, in all targets, the half-light radius, estimated from the \prism total flux, falls between the first and second annuli. For each annulus, the SFH was derived by summing the SFR in each time bin over all pixels within the ring. 
The resulting radial SFHs are shown in Fig.~\ref{fig:sfh_radius}. 

In GS10578, the SFH shows clear radial variations: the central regions form stars earlier and more intensely, reaching SFRs up to $\sim100$~M$_{\odot}\,\mathrm{yr}^{-1}$, consistent with an inside-out growth. A sharp decline in SFR is observed at all radii $\sim 200$~Myr ago, indicating a galaxy-wide suppression of star formation.
In GS19293, the SFR remained roughly constant across cosmic time. At $z = 3-4$, the inner and outer regions show mild variations (likely consistent with an inside-out growth), converging into a final burst of star formation at the epoch of observation across all radii.
In GS20936, the central region exhibits higher SFR during the first Gyr compared to the outer regions, consistent with inside-out growth. The central SFR then declines, while the outer regions remain approximately constant, followed by an increase in SFR at the epoch of observation across all radii.
In GN20, the SFR remains relatively constant across all radii for $\sim1$ Gyr, then rises to a peak (up to $\sim 1000$~M$_{\odot}\,\mathrm{yr^{-1}}$
) around 200 Myr before the epoch of observation ($z\sim5$), followed by a further burst at the time of observation.
In the case of the two high-$z$ starbursts, EGSY8p7 and COS12306, the SFH curves are similar at all radii, with the highest SFRs occurring in the central region at the time of observation. In the case of COS12306, we identified the southern clump as the nucleus (see Fig. \ref{fig:12306-COS}); therefore the annuli follow the galaxy’s elongated morphology. However, the fact that they are associated with an earlier epoch and have a compact morphology makes it more difficult to trace radial variations.  

To summarize, the four lower-redshift ($z\sim3$) AGN-host galaxies display signatures consistent with an inside-out growth scenario. Such trends are less evident in our starburst galaxies $z > 5$, 
likely because they are still in an early phase of assembly and have not yet developed a well-defined nucleus. Alternatively, this may be due to limitations imposed by angular resolution. 

\begin{figure*}
    \centering
    \includegraphics[width=1\linewidth]{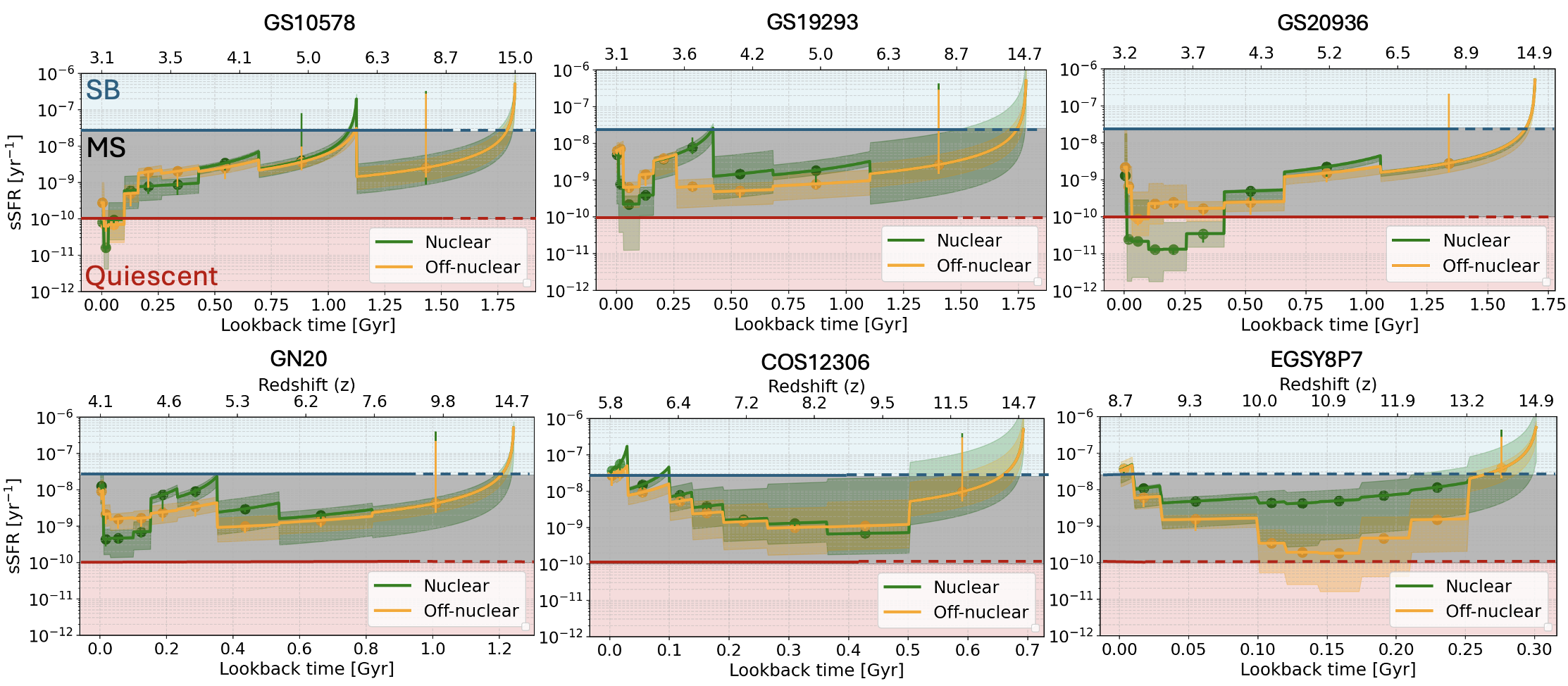}
    \caption{Specific star formation rate (i.e. SFR/M$_\star$) as a function of the lookback time and redshift. Blue shaded area represents the starburst (SB) regime, grey one represents the main sequence (MS), and  red the quiescent regime. Green sSFR curves refer to nuclear region, while yellow curves refer to the outer region. The dots represent the sSFR at the average time of the bin, while the vertical lines associated with each point represent the minimum and maximum sSFR over the time interval, assuming in the nonparametric model that the SFR is constant within each time bin. The solid horizontal lines indicate the region at $z\lesssim 9$, where the redshift-independent and -dependent definitions provide broadly consistent classifications for starbursts, main sequence, and quiescent systems.} 

    \label{fig:ssfr}
\end{figure*}

\subsubsection{Nuclear and off-nuclear sSFR evolution: main-sequence and off-main-sequence phases}\label{sec:nucleus_vs_outer}
While the SFR provides an absolute measure of star formation, the sSFR allows us to account for the impact on the total stellar mass and thus better trace the relative growth of galaxies and establish if the galaxy is growing in a steady state on the main sequence, or it is taking excursions from it. From here on, we classify the galaxies in our sample according to their sSFR into three regimes: starburst, main-sequence, and quiescent systems.  We define as being in a quiescent phase the galaxies with $\mathrm{sSFR}/\mathrm{yr^{-1}} \leq 10^{-10}\,$, following \cite{Franx2008} and \cite{Kurinchi-Vendhan2024}. For simplicity, we preferred a redshift-independent definition of quiescence, noting that our results do not change when adopting $\mathrm{sSFR} < 0.2/t_{\rm H}(z)$ where $t_{\rm H}$ is the Hubble time (\citealt{Carnall2023}).
Following the criteria established by \citet{Caputi2017,Caputi2021}, we define main-sequence galaxies as those with 
$  {-10} < \log(\mathrm{sSFR}/\mathrm{yr^{-1}}) <  -7.60$
and starburst galaxies as those with 
$\log(\mathrm{sSFR}/\mathrm{yr^{-1}}) > -7.60$.

We focused on a central region, corresponding to approximately $2\sigma$ of the PSF ($r = 0.2''$), and an outer region ($r > 0.2''$) in order to investigate how the sSFR evolves in terms of its position relative to the star-forming main sequence, and whether the galaxies exhibit phases of quenching or starburst activity.
Figure \ref{fig:ssfr} shows the sSFR as a function of lookback time in the central region (green curve) and outside (yellow curve). To compute the sSFR, we divided the total SFR by the total stellar mass within the central region and in the outer region. 
Note that within the same time bin, the trend of the sSFR function decreases as the lookback time decreases, due to the construction of the SFH, which assumes a constant SFR in each bin while the mass increases. Therefore, for each temporal bin, the sSFR varies between the sSFR at the beginning of the bin and the sSFR at the end of the time bin (vertical lines). The shaded area instead represents the uncertainties on the sSFR for each bin.

Figure \ref{fig:ssfr} shows that all the galaxies in our sample, from the onset of star formation onwards, remain on the main sequence most of their lifetime both in the central  and in the outern regions. GS10578 and GS20936 reach the quenching phase  at z$\sim$3.2 and z$\sim$3.7, respectively. In GS20936, the central region undergoes complete quenching, while the outer regions, despite a decrease in sSFR, do not reach quenching. This result indicates that the inner regions experienced a more powerful quenching compared to their outer regions (see also \citealt{Haryana2025}). Instead, in GS10578, the sSFR reaches quenching levels both within the nuclear and in the outer regions. 
It is interesting to note that in both GS10578 and GS20936 we observe a subsequent rise in sSFR that brings the galaxies back toward the main sequence. In GS10578, this occurs only outside the nucleus. 
In contrast, GS20936 shows clear evidence of rejuvenation, returning to the main sequence after a quenching phase both in the nuclear and off-nuclear regions. 

The sources COS12306 and EGSY8p7, with the highest redshift and lowest stellar masses in our sample, after spending around 200–500 Myr on the main sequence, transition into a starburst phase at the time of observation. In these cases, the sSFR in the nuclear and off-nuclear regions do not differ significantly. 

In GS19293 and GN20, the trends in sSFR between the nuclear and outer regions are very similar: both remained on the main sequence, albeit with a wide scatter.  

Overall, we found two galaxies that reach the quenching phase (GS10578 and GS20936), mostly in the nuclear region. The lowest sSFR is found in GS20936, reaching values of $10^{-2} \, \mathrm{Gyr^{-1}}$ in the nuclear region. 
The other two AGN galaxies, GN20 and GS19293, 
remained on the main sequence (even with large variations, likely reflecting the scatter observed at high redshift in the main sequence; \citealt{Calabro2024}). 
Finally, the two starburst galaxies (COS12306 and EGSY8p7) spent the majority of their life on the main sequence, reaching the starburst phase only in the last few 10s~Myr.

\subsection{Outflows as the cause of quenching phases
}\label{subsec:outflows}

\begin{table*}
\caption{Spatially resolved physical properties of the targets from this work and from literature. }
\tabcolsep 4.4pt 
\label{tab:properties_outflow}
\centering\small
\begin{tabular}{lcccccccc}
\hline\hline
Target &
$M_{\star,\mathrm{res}}^{(1)}$ &
log$_{10}$\,SFR$_{10}^{(2)}$ &
log$_{10}$\,SFR$_{100}^{(3)}$ &
log$_{10}$\,SFR$_{\mathrm{line}}^{(4)}$ &
$\dot{M}_{\mathrm{out}}^{(5)}$ &
$\eta^{(6)}$ &
Activity &
SFH features \\
&
[$10^{10}M_\odot$] &
[$M_\odot\,\mathrm{yr}^{-1}$] &
[$M_\odot\,\mathrm{yr}^{-1}$] &
[$M_\odot\,\mathrm{yr}^{-1}$] &
[$M_\odot\,\mathrm{yr}^{-1}$] &
&
&
\\
\hline

GS10578$^{a}$ &
$38^{+8}_{-8}$ &
1.9$^{+0.9}_{-1}$ &
1.5$^{+0.5}_{-0.7}$ &
2.0$^{+0.2}_{-0.2}$ &
$27^{+0.9}_{-0.9}$ &
$0.84^{+3}_{-0.5}$ &
AGN (X-ray+BPT) &
quench \\

GS19293$^{a}$ &
$15^{+5}_{-5}$ &
$2.95^{+0.6}_{-0.8}$ &
$2.46^{+0.5}_{-0.5}$ &
$1.52^{+0.3}_{-0.2}$ &
$0.11^{+0.017}_{-0.014}$ &
$0.0004^{+0.001}_{-0.0002}$ &
AGN (BPT) &
MS \\

GS20936$^{a}$ &
$49^{+9}_{-10}$ &
$2.9^{+1.2}_{-0.9}$ &
$2.1^{+0.8}_{-0.6}$ &
$2.0^{+0.3}_{-0.3}$ &
$18.2^{+0.9}_{-0.9}$ &
$0.13^{+0.6}_{-0.07}$ &
AGN (X-ray+BPT) &
quench + rej
 \\

GN20$^{b,}$ &
$97^{+10}_{-25}$ &
$3.95^{+0.4}_{-0.7}$ &
$3.3^{+0.3}_{-0.5}$ &
$2.1^{+0.2}_{-0.2}$ &
$0.7^{+0.3}_{-0.3}$ &
$0.0003^{+0.001}_{-0.0001}$ &
AGN (Broad lines) &
MS \\

COS12306$^{d}$ &
$0.29^{+0.04}_{-0.06}$ &
$1.9^{+0.2}_{-0.2}$ &
$1.4^{+0.1}_{-0.1}$ &
$2.2^{+0.5}_{-0.5}$ &
$50^{+30}_{-20}$ &
$2^{+0.5}_{-0.36}$ &
Starburst &
rising SFR \\

EGSY8P7$^{c,d}$ &
$0.34^{+0.06}_{-0.09}$ &
$2.0^{+0.3}_{-0.2}$ &
$1.2^{+0.1}_{-0.2}$ &
$1.7^{+0.5}_{-0.6}$ $(2.3^{+1.9}_{-1.9})$ &
$7.8^{+0.7}_{-0.7}$ $(132^{+1450}_{-107})$ &
$0.45^{+0.2}_{-0.3}$ &
Starburst &
rising SFR \\

\hline
\end{tabular}

\tablefoot{Columns (1), (2) and (3) are derived from resolved SED analysis. (4) refers to the  literature SFR computed via recombination lines. For GN20 we report the SFR computed from \cite{Bik2024} with Pa$\alpha$.  (5) is the mass outflow rate of the  ionized phase computed through $\left[\rm O \, \textsc{iii}\right]$ emission line. (6) is the mass loading factor $\dot{M}_{\mathrm{out}}/\mathrm{SFR}_{100}$. 
Letters next to target names indicate literature references of the $\dot{M}_{\mathrm{out}}$: (a) \cite{Venturi2025}; (b) \cite{Ubler2024} 
; (c) \cite{Zamora2025}; (d) \cite{RodriguezDelPino2026}.
For EGSY8P7 the values in (4) and (5) are from \cite{Zamora2025}, while values in parentheses are from \cite{RodriguezDelPino2026}. (7) and (8) are activity and SFH-feature labels, inferred from this study and literature classifications; they should be interpreted as heuristic rather than definitive.}
\end{table*}


In this section we provide insights into the quenching mechanism by linking the SFH derived in the previous section with the properties of the ionized outflows derived in previous works (\citealt{Ubler2024,Venturi2025, Zamora2025,RodriguezDelPino2026}). However, we stress that the interpretation of these results is subject to several limitations, including uncertainties in the derivation of outflow properties and the mismatch in timescales between SFHs and outflow measurements; these are discussed in detail in Appendix~\ref{app:Caveats}.


A commonly used diagnostic to investigate the outflow impact on the SF and its potential quenching is the mass loading factor ($\eta$), defined as the ratio of the mass outflow rate ($\dot{M}_{\mathrm{out}}$) to the SFR (e.g. \citealt{RodriguezDelPino2026}). 
This quantity provides a first-order measure of the balance between gas consumption through star formation and gas depletion driven by outflows. In particular, $\eta \gtrsim 1$ indicates that the gas is being removed more rapidly than it is converted into stars, suggesting that outflows can efficiently deplete the gas reservoir and potentially lead to quenching if  maintained at the current rates for a sufficient time. Conversely, $\eta \lesssim 1$ implies that star formation dominates over gas removal, allowing the galaxy to sustain its growth despite the presence of outflows. 
For each galaxy, we estimate $\eta$ using the ionized mass outflow rates reported in Table \ref{tab:properties_outflow}  and the global SFR derived from the resolved analysis averaged over the last 100~Myr, a timescale roughly similar to the AGN duty cycle. Across the sample, ionized outflows are detected on spatial scales of $\leq 1$ to $\sim 3$~kpc, with mass outflow rates spanning $\sim 0.1$ to $\gtrsim 100$~\Msunyr. The outflow properties adopted in this work and the references from which they were taken are summarised in Table \ref{tab:properties_outflow}.

In the most massive systems, GS10578 and GS20936, which also show the strongest X-ray emission (> $L_{\rm 2-10~keV} \sim 8 \times 10^{44}\ \mathrm{erg\ s^{-1}}$; \citealp{Luo2017_cat}), the inferred ionized mass-loading factors are the highest within our AGN sub-sample (see Table \ref{tab:properties_outflow}). Moreover, for GS10578, \cite{DEugenio2024} reported a substantial neutral mass outflow rate (30–300 $M_\odot\,\mathrm{yr}^{-1}$), implying a total mass-loading factor of $\eta>>1$. For these two systems the associated powerful AGN-driven outflows are concentrated in the nuclear regions (<3~kpc, \citealt{DEugenio2024, Venturi2025}). 
In these galaxies, the SFHs indicate a significant decline in star formation in the last 100--300 Myr, particularly in the nuclear regions (Fig. \ref{fig:ssfr}), suggesting that previous outflow episodes of similar intensity may have played a role in regulating star formation activity.  Although we cannot probe directly past outflow events, the fact that these two galaxies assembled most of their stellar mass prior to the onset of the quenching period (see Fig. \ref{fig:resolved_unresolved_mass}) suggests that, given the expected BH-galaxy co-evolution, they already hosted AGN capable of driving outflows of comparable power during that epoch. Therefore, outflows seem to have had a significant role in shaping these systems.

The other two AGN, GN20 and GS19293, show weaker X-ray emission and exhibit comparatively low mass loading factors (<0.0005). Moreover, their SFHs do not show strong evidence for a significant decline in star formation at any epoch, and they have remained on the main sequence along their lifetimes (Fig. \ref{fig:ssfr}). Therefore, the currently observed outflows are unlikely to be associated with recurrent episodes capable of dominating the evolution of the star formation activity in the host galaxies. 

The two $z>5$ starburst systems, EGSY8P7 and COS12306, stand apart, exhibiting a markedly different behavior. Both show substantial outflow rates and relatively high mass loading factors, yet their SFHs do not display either clear signatures of long-term suppression nor steady growth on the main sequence. In this case, the observed outflows seem associated with the recently enhanced star formation activity rather than sustained feedback capable of significantly suppressing star formation or maintaining the growth on a steady mode. Indeed, SF-driven outflows were likely weaker in the past: the SFH inferred from Fig. \ref{fig:resolved_unresolved} indicates SFRs more than ten times lower just $\simeq$50 Myr ago.


In summary, although it remains challenging to directly link outflow properties to the suppression of star formation, the diversity of behaviours in our sample allows us to outline possible evolutionary scenarios. The two most massive galaxies hosting X-ray AGN (GS10578 and GS20936) show evidence of past suppression of star formation, and their powerful outflows are consistent with ongoing feedback. If such outflows are observed at the present epoch, it is plausible that similar episodes occurred in the past, potentially contributing to the quenching phases inferred from the SFHs at lookback times of a few hundred Myr. This is supported also by the large masses built up at the redshifts when the quenching phase began (Fig. \ref{fig:resolved_unresolved_mass}), suggesting that the BHs could have been active at those epochs, possibly generating  gas outflows that may have played a role in shaping their evolutionary paths. In contrast, the two starburst systems display similarly strong outflows but no clear signatures of quenching, indicating that these outflows, likely driven by recent star formation episodes, may not yet have had a significant impact on the host galaxy evolution. GN20 and GS19293 appear to represent intermediate cases, where AGN activity may be present but has not yet had a significant impact on the global star formation.

\subsection{The role of merging in `rejuvenating' galaxies}\label{sec:rejuvenation}

The two targets that experienced quenching episodes, GS10578 and GS20936, also show a recent enhancement in their star formation activity: their SFRs reach $10^{2}-10^3$~\Msunyr in the last 10~Myr. Following \citet{Harrold2026}, we define a `rejuvenated' galaxy as one that transitions from a quiescent phase back to the main sequence. Under this definition, only GS20936 qualifies as a rejuvenated system. Although GS10578 also shows renewed star formation, this is confined to the outer regions and does not correspond to a global return to the main sequence. 
In the next two subsections, we discuss the possible origin of the recent increase of SFR in our two systems and use TNG simulations to gain insight into the mechanisms driving it.

\subsubsection{Merger-driven rejuvenation scenarios}

The morphology and spatial distribution of the recent star formation provide important clues on the physical processes responsible for the rejuvenation phase.
Both GS10578 and GS20936 show enhanced star formation in extended, clumpy off-nuclear regions. 
Several mechanisms may be responsible for this, including mergers, cold gas accretion from cosmic web, or a combination of both. 
The presence of blue clumps in the outer regions of GS10578, as revealed in Fig.~\ref{fig:three_colors} (see also \citealt{DEugenio2024}), suggests that the enhanced SFR may be linked to minor-merger activity, which primarily deposits material in galaxy outskirts (see Fig.~\ref{fig:ssfr}) and leads to more localized and subtle changes in SFR (e.g. \citealt{Lambas2012,Jackson2022}).
For GS20936, NIRCam imaging (Fig. \ref{fig:rej20936}) reveals a major merger configuration, with filamentary structures connecting it to a nearby massive companion ($z_{\rm ph} = 3.18$, \citealt{Hainline2023}) at a projected distance of $\sim20$~kpc. This morphology suggests that the galaxy may have already experienced a pericentric passage, during which a high amount of gas may have been funnelled into GS20936, triggering a rejuvenation episode  both in the inner and outer regions of the host galaxy.

Linking merger activity to enhanced star formation and clumpy morphologies is a well-established picture across cosmic time, with similar signatures also observed in local interacting ultra luminous infrared galaxies (ULIRGs; e.g. \citealt{Arribas2004,GarciaMarin2009,PiquerasLopez2013,Zaragoza-Cardiel2018,Larson2020}).
The merger interpretation is further supported by cosmological simulations (e.g. \citealt{Naab2009, Oser2012,  Wellons2015, Rodriguez-Gomez2016}). 

Although both GS10578 and GS20936 show evidence for recent star formation enhancement, the duration and future evolution of this phase remain uncertain. The increase in SFR is observed only within the last few tens~Myr, and may therefore correspond to the early stages of a rejuvenation episode associated with the ongoing interactions. Consequently, it is not yet possible to robustly estimate the total stellar mass that will ultimately be formed during this phase; within the timescales currently probed, we infer only a modest increase in stellar mass of $\approx 1$\% in both systems. Numerical simulations and observational studies suggest that rejuvenation episodes can persist for several hundred Myr up to $\sim 1$ Gyr, depending on the gas supply and merger history (\citealt{Pandya2017}). In this context, the two galaxies analysed here may represent systems observed during the early stages of a rejuvenation event.

\subsubsection{Comparison with TNG simulations}

To further investigate the physical origin of rejuvenation episodes, we compare our observational results with galaxies extracted from the TNG300 cosmological simulation (\citealt{Nelson2018}), based on the state-of-the-art code Arepo described by \cite{Springel2010} and \cite{Weinberger2020}. Our goal is twofold: first, we investigated the presence and incidence of 
mock galaxies undergoing rejuvenation episodes similar to GS20936; 
second, we traced the history of their progenitors to explore the physical mechanisms driving both the initial quenching and the subsequent resumption of star formation, and to assess whether these mechanisms are consistent with our observational results. 

\begin{figure}
    \centering
    \includegraphics[width=1\linewidth]{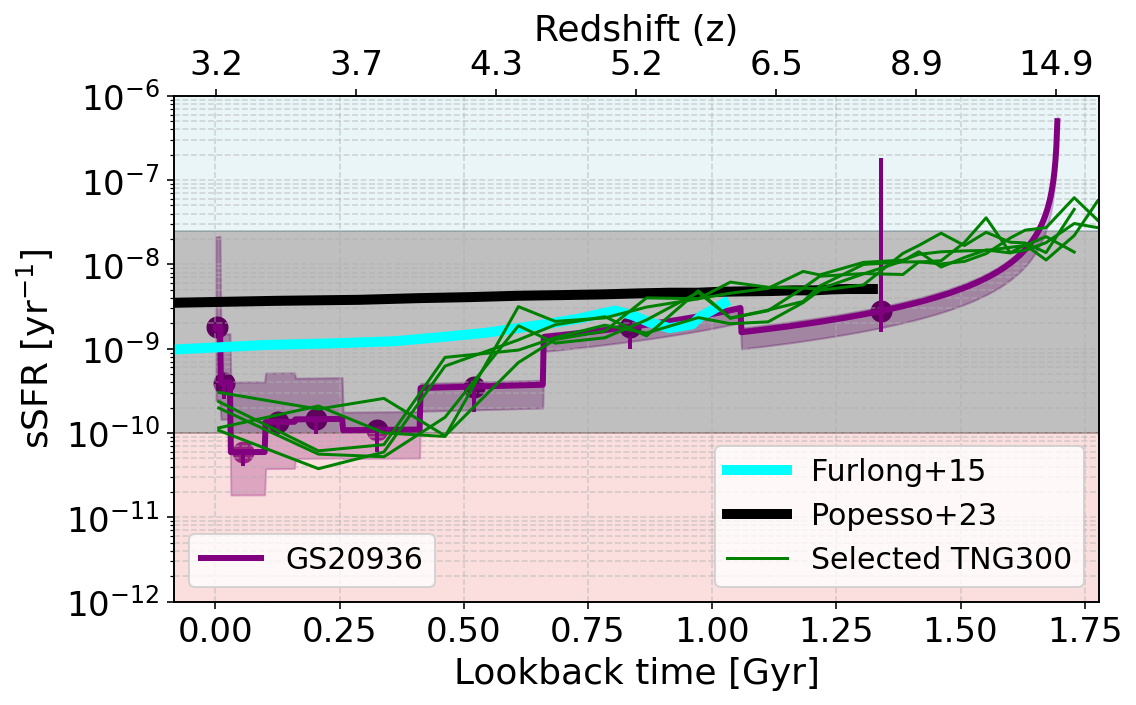}
    \caption{Total sSFR history of GS20936 estimated by summing individual spaxels (purple curve) with associated error (purple shaded area). Green curves are the six selected rejuvenated galaxies from TNG300. Cyan and black curves represent the main sequence from EAGLE simulations (\citealt{Furlong2015}) and from \cite{Popesso2023}.}
    \label{fig:ssfrTNG300}
\end{figure}

We searched for objects in the TNG300 simulations that 
are sufficiently massive at $z=5$ and exhibit a SFH similar to GS20936. 
To do so, we 
filtered galaxies with $M_{\star} > 5 \times 10^{10}\, \mathrm{M}_\odot$ at $z = 5$, similar to GS20936 (Fig.~\ref{fig:resolved_unresolved_mass}). This search led to the selection of 119 mock galaxies.
Among them, 
we selected only those that entered a quenched (sSFR $ < 10^{-10}~\mathrm{yr^{-1}}$) phase at $z>3$ and subsequently underwent a rejuvenation phase, according to the definition in Sect.~\ref{sec:rejuvenation}.
This selection leads to six galaxies, corresponding to  5\% of the parent sample of TNG300 galaxies selected based on stellar mass.
In Appendix \ref{sec:Snapshot} we show the sSFR evolution of the six selected rejuvenated galaxies together with all the 119 massive galaxies  (top-left panels in Figs.~\ref{fig:ID1625}-\ref{fig:ID45033}).
We compared in Fig. \ref{fig:ssfrTNG300} the sSFR history of GS20936, calculated by summing pixel by pixel from the total aperture, alongside that of the rejuvenated galaxies identified in TNG300 simulations. We note that the overall behaviour of the sSFR curves is very similar 
although in GS20936 there is a steeper increase at the redshift of the source. 
For reference, we also compare our results with the main-sequence star-forming galaxies at $0<z<6$ from observational measurements  (\citealt{Popesso2023}) and the EAGLE cosmological simulations (\citealt{Furlong2015}), finding that both the selected TNG galaxies and GS20936 exhibit a steeper decline towards the quenched phase.

We then followed the growth of the black holes hosted in the six mock galaxies to 
assess whether their quenching is consistent with AGN feedback, and specifically whether the central black hole was actively accreting at the time when star formation was suppressed. 
We found that from $z=6$ down to $z=4$, the black hole undergoes a phase of very intense accretion (see top right panels in Figs.~\ref{fig:ID1625}-\ref{fig:ID45033}) 
; therefore, it is conceivable that the high activity of the AGN may have heated up the gas or triggered outflows that suppressed star formation in the past. 

We further investigated the temporal evolution of these galaxies in order to assess whether their rejuvenation episodes are associated with major or minor merger events. In Fig.~\ref{fig:ID1625}, we show five temporal snapshots of both stellar mass (red) and gas mass distribution (blue) of one of the six selected TNG galaxies. 
We find that at $z \sim 7$ (snapshot E), the galaxy is compact and appears to accrete gas through filamentary structures. Before quenching (snapshot D), the system does not show clear signatures of ongoing mergers. The galaxy then undergoes a quenching phase, while in the more recent snapshots (A, B, and C), after the quenching, it enters a merging stage that coincides with an increase in the SFR, consistent with the rejuvenation episode. This evolutionary pathway may resemble what we are observing in GS20936 and, partially, in GS10578.
The sSFR evolution curves (and snapshots) for the other five TNG galaxies selected as rejuvenated systems are presented in Appendix \ref{sec:Snapshot}. Most of the galaxies show merger signatures in the most recent snapshots after quenching. 

Overall, the comparison with TNG300 suggests that rejuvenation episodes similar to that observed in GS20936 are relatively rare among massive galaxies at $z > 3$, occurring in only a few percent of the simulated population. The rejuvenation phase is associated with merger activity occurring after a previous quenching episode, supporting a scenario in which galaxy interactions contribute to re-supplying gas and reactivating star formation. The similarities between the simulated systems and our observations indicate that GS20936, and possibly GS10578, may represent galaxies observed during a transient phase of renewed growth following an earlier quenching event.


\section{Conclusions}\label{sec:Conclusions}
In this work, we used \textit{JWST}/NIRSpec IFS \prism observations to investigate the continuum emission and study the stellar properties of six AGN and starburst galaxies at $3<z<9$, deriving stellar masses and SFHs from both spatially resolved and integrated spectra. We compared the SFHs with the ionised outflows properties inferred from the high spectral resolution (R~2700) NIRSpec obsevations of our sample \citep[][]{Ubler2024,Zamora2025,Venturi2025,RodriguezDelPino2026}, in order to  assess the possible role of feedback shaping the SFHs.
The main findings can be summarized as follows:

\begin{itemize}

\item  By comparing the spectra of individual spaxels with spatially integrated spectra, we quantify the so-called outshining effect, where stellar masses obtained from integrated spectra are systematically underestimated compared to those derived from resolved measurements. We found that the magnitude of this effect increases with sSFR, reaching up to $\sim$0.75 dex. 
This highlights the importance of spatially resolved analyses for robust stellar mass estimates at high redshift. 

\item  The SFHs reveal a diversity of evolutionary pathways. Massive galaxies hosting AGN (GS10578, GS20936, GS19293, GN20) show more complex and non-monotonic histories, while lower-mass systems at earlier redshift (COS12306, EGSY8P7) exhibit smoother, steadily rising SFHs with no clear signatures of quenching. 

\item Two galaxies feature SFHs with short periods of quiescence  (GS10578 and GS20936), with GS20936 showing the lowest sSFR in our sample ($\sim 10^{-11}~ \mathrm{yr^{-1}}$ in the nuclear region). GN20 and GS19293 display fluctuations in their SFHs, but broadly remained in the main sequence over their lifetimes. The two starburst galaxies (COS12306 and EGSY8p7) reached the starburst phase only in the last few million years.


\item Our study suggests a diversity of evolutionary stages linking star formation, AGN activity, and gas outflows. The two high-stellar mass galaxies (GS10578 and GS20936) with clear AGN features show evidence of a previous quenching phase and have strong outflows that may have contributed to the quenching. In contrast, the two starburst systems also exhibit strong outflows, but these are more likely driven by their recent enhancement in star formation activity, with no indication of a prior quenching phase and no evidence of AGN-driven feedback. Finally, GN20 and GS19293 appear to represent intermediated cases, where AGN activity may be emerging but has not yet significantly affected the global star formation properties of the galaxies due to the weaker outflows.


\item We studied in more detail the case of GS20936, which after a quenching episode  has experienced a rejuvenation phase with a significant SFR enhancement. In order to assess how typical this pattern is, we used TNG300 simulations, finding that $\sim 5\%$ of massive galaxies ($M_{\star} > 5\times10^{10}\,M_{\odot}$ at $z=5$) exhibit evolutionary trends similar to those observed in GS20936.   
In these simulated GS20936 analogues, we often identify merger signatures that may have contributed to the rejuvenation of star formation by supplying fresh gas to the systems and interrupting quenching processes. 
\end{itemize}

This work highlights the importance of spatially resolved observations of galaxies at $3 < z < 9$, providing a clearer picture of the role of AGN-driven feedback in shaping their evolution. Our results suggest that massive galaxies at $z \sim 3$ experienced quenching episodes likely following powerful outflow events (similar to those currently observed in their ionized gas phase); instead, in more distant ($z > 5$) starburst galaxies, the outflows originated from intense star formation activity and have not yet led to quenching. 
Although our study is based on a limited sample of only six systems, the results reveal a wide diversity of evolutionary pathways at early cosmic epochs. Extending similar spatially resolved analyses to larger and more representative samples, including not only AGN hosts and starbursts but also main-sequence galaxies, will be essential to establish the relative role of feedback, mergers, and gas accretion in shaping galaxy evolution across cosmic time.


\begin{acknowledgements}
The full acknowledgements are available in Appendix \ref{app:ack}.
\end{acknowledgements}

\bibliographystyle{aa}
\bibliography{references} 

\appendix

\section{Acknowledgements}\label{app:ack}

LU, MP, SA, and BRP acknowledge support from the research project PID2021-127718NB-I00 of the Spanish Ministry of Science and Innovation/State Agency of Research (MCIN/AEI/10.13039/501100011033).
GC acknowledges the support of the INAF Large Grant 2022 "The metal circle: a new sharp view of the baryon cycle up to Cosmic Dawn with the latest generation IFU facilities". GC acknowledges support from PRIN-MUR project "PROMETEUS" (202223XPZM), the INAF Large Grant 2022 “The metal circle: a new sharp view of the baryon cycle up to Cosmic Dawn with the latest generation IFU facilities” and INAF Large Grant 2022 "Dual and binary SMBH in the multi-messenger era".  
AJB acknowledges funding from the “FirstGalaxies” Advanced Grant from the European Research Council (ERC) under the European Union’s Horizon 2020 research and innovation program (Grant agreement No. 789056).
GV, SZ, SC acknowledges support from European Union’s HE ERC Starting Grant No. 101040227 - WINGS.
IL is supported by the European Research Council (ERC) under the European Union’s Horizon 2020 research and innovation programme (DistantDust, Grant agreement No. 101117541).
H\"U acknowledges support by the Max Planck Society through the Lise Meitner Excellence Program. H\"U acknowledges funding by the European Union (ERC APEX, 101164796). Views and opinions expressed are however those of the authors only and do not necessarily reflect those of the European Union or the European Research Council Executive Agency. Neither the European Union nor the granting authority can be held responsible for them.
B.R.P acknowledges support from grant PID2024-158856NA-I00 funded by Spanish Ministerio de Ciencia e Innovación MCIN/AEI/10.13039/501100011033 and by “ERDF A way of making Europe

\section{PSF modelling}\label{app:A}
Given the uncertainties in the PSF measurements obtained using different methods in \cite{DEugenio2024} and \cite{Jones2026}, we selected a bright standard star observed with NIRSpec \prism (PID: 1219, PI: N.~L\"utzgendorf), Gaia EDR3 3725409861811527680, to characterize the PSF. To this end, we modeled each  slice of the JWST IFU data cubes with a two-dimensional Gaussian. The functional form of the model is

\begin{equation}
g(x,y) = \mathrm{offset} + A \, \exp \Bigg\{ -\frac{1}{2} \Big[ \left( \frac{x' - x_0}{\sigma_\mathrm{maj}} \right)^2 + \left( \frac{y' - y_0}{\sigma_\mathrm{min}} \right)^2 \Big] \Bigg\},
\end{equation}

where $(x_0, y_0)$ is the center of the Gaussian, $A$ is its amplitude, $\sigma_\mathrm{maj}$ and $\sigma_\mathrm{min}$ are the major and minor axis standard deviations, $\theta$ is the position angle of the major axis with respect to the detector axes, and \textit{offset} accounts for the background level. The rotated coordinates $(x',y')$ are defined as

\begin{align}
x' &= (x-x_0)\cos\theta + (y-y_0)\sin\theta,\\
y' &= -(x-x_0)\sin\theta + (y-y_0)\cos\theta.
\end{align}


The Gaussian fit was performed independently for each spectral slice using a non-linear least-squares minimization (Levenberg-Marquardt) implemented with \texttt{scipy.optimize.curve\_fit}. 


We find that the $\sigma$ values estimated by \cite{DEugenio2024} are systematically smaller than our measurements, while the geometric mean reported by \cite{Jones2026} 
shows perfect agreement. 
Interestingly, the centre of the PSF exhibits a wavelength dependence as well: for wavelengths below $\sim 2\,\mu$m the offset varies roughly in the same direction as $\theta$, whereas at longer wavelengths it shifts along an angle of approximately $120^\circ$ relative to $\theta$ 
(see also e.g. \citealt{Pascalau2026}).


\section{Photometric factor NIRCam - NIRSpec}\label{app:B}
We rescale the \prism spectrum to the NIRCam photometry by fitting a polynomial correction factor within our SED modeling (\citealt{Wang2025,DEugenio2024}). 
We used the JADES photometric catalog\footnote{ 
\texttt{hlsp\_jades\_jwst\_nircam\_goods-s\_photometry\_v5.0\_catalog.fits}}
to extract NIRCam fluxes in the F090W, F115W, F150W, F200W, F277W, F356W, and F444W filters in an aperture of 0.5" (CIRC6, JADES Data Release 5, \citealt{Johnson2026, Robertson2026}). 
For each galaxy, we derived synthetic fluxes from the NIRSpec \prism spectra by convolving the observed spectra with the corresponding NIRCam filter transmission curves. 
For a given filter, the effective flux density was computed as the transmission-weighted mean of the spectral flux density across the bandpass,
\begin{equation}
\langle F_{\lambda} \rangle = 
\frac{\int F_{\lambda}(\lambda)\,T(\lambda)\,d\lambda}
{\int T(\lambda)\,d\lambda},
\end{equation}
where $F_{\lambda}(\lambda)$ is the observed spectral flux density and $T(\lambda)$ is the filter transmission curve. 
The resulting effective fluxes were converted to $F_{\nu}$ using the effective wavelength of each filter.

\begin{figure}
    \centering
    \includegraphics[width=1\linewidth]{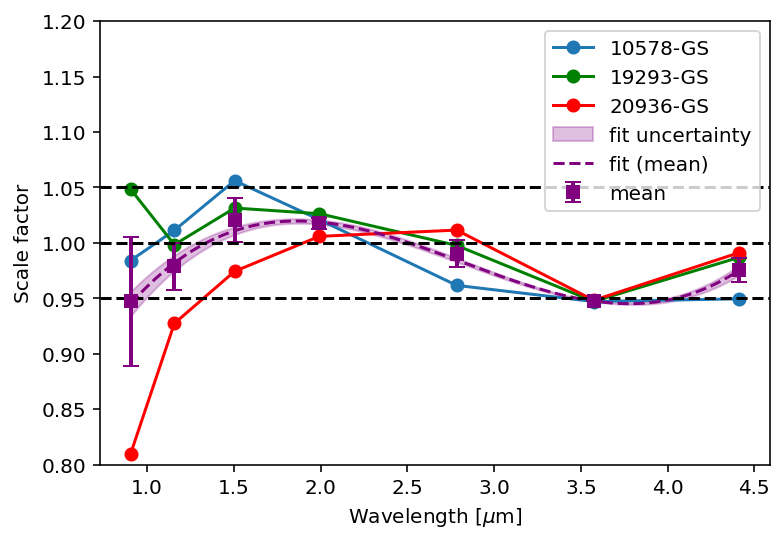}
    \caption{Scale factors required to match the NIRSpec \prism spectroscopy with the NIRCam photometry. The blue, red, and green points and lines correspond to the galaxies GS10578, GS20936, and GS19293, respectively. Purple squares indicate the mean scaling factor, with error bars representing the error associated among the three galaxies. The shaded region shows the $1\sigma$ uncertainty on the third-degree polynomial fit, computed from the covariance matrix of the fit coefficients.
}
    \label{fig:scalefactor}
\end{figure}

We then compared the synthetic NIRSpec fluxes within an aperture of 0.5" with the observed NIRCam photometry and computed, for each filter, a scaling factor between spectroscopy and imaging for the galaxies in the GOODS-S field included in our sample: GS10578, GS19293, and GS20936. We then considered the mean value of the three galaxies in each filter and fitted a third degree polynomial. We then used the best-fit polynomial for the \prism spectrum to perform a flux calibration to the \prism data shown in Fig~\ref{fig:scalefactor}.
All three galaxies exhibit a similar wavelength-dependent behaviour of the scaling factor. 
In particular, the scaling factor oscillates at the 5--10\% level, increases in the wavelength range 1--1.5~$\mu$m, decreases between 1.5 and 3.5~$\mu$m, and rises again at longer wavelengths. 

\section{SED fitting parameter space}\label{app:sedfitting_parameter}

Table \ref{tab:setup}  summarizes the parameter space explored during the Prospector fitting procedure for the non parametric, parametric and parametric+burst models.

\begin{table}[ht]
\centering
\caption{Summary of SED fitting parameter space and assumptions explored using \label{tab:setup}\texttt{Prospector}}

\begin{tabular}{p{0.43\columnwidth} p{0.43\columnwidth}}
\toprule
Free Parameter & Range \\
\midrule

SFH bins (lookback time in Myrs) 
& $0$-$10$,$0$-$10$,$20$-$100$, bin4-9 log spaced up to $z=15$ \\

SFH prior 
& \texttt{continuity\_sfh} with a flat prior $-2 < \log \mathrm{SFR}_{i}/\mathrm{SFR}_{i+1} < 2$\\

t$_{age}$ and time of the burst (\textit{tburst}) & between time of observation and z=15 \\

Fraction of mass formed during burst & $0 < fburst < 0.8$ \\

Dust optical depth (\citealt{Charlot2000})
& $0 < \tau_{\mathrm{dust}} < 10.0$ \\

Stellar mass 
& $7.0 < \log_{10}(M_\ast/M_\odot) < 12.0$ \\


\multicolumn{2}{p{\columnwidth}}{MILES stellar library \& MIST stellar isochrones} \\

Stellar metallicity 
& $-2.5 < \log_{10}(Z/Z_\odot) < 0.5$\\



Ionization parameter 
& $-4.0 < \log_{10}U < -1.0$ \\

\midrule
Other parameters & Type \\
\midrule

IMF 
& \cite{Kroupa2001} ($0.08\,M_\odot$ - $120\,M_\odot$) \\

Spectral smoothing (\texttt{smooth\_type}) 
& LSF \\

Dynesty sampling
& \texttt{nlive\_init}=500, \texttt{nested\_method}=rwalk, \texttt{maxcall}=$7\times10^{6}$ \\

\bottomrule
\end{tabular}
\end{table}

\section{Stellar mass and extinction maps}\label{app:C}

\begin{figure*}
    \centering
    \includegraphics[width=1\linewidth]{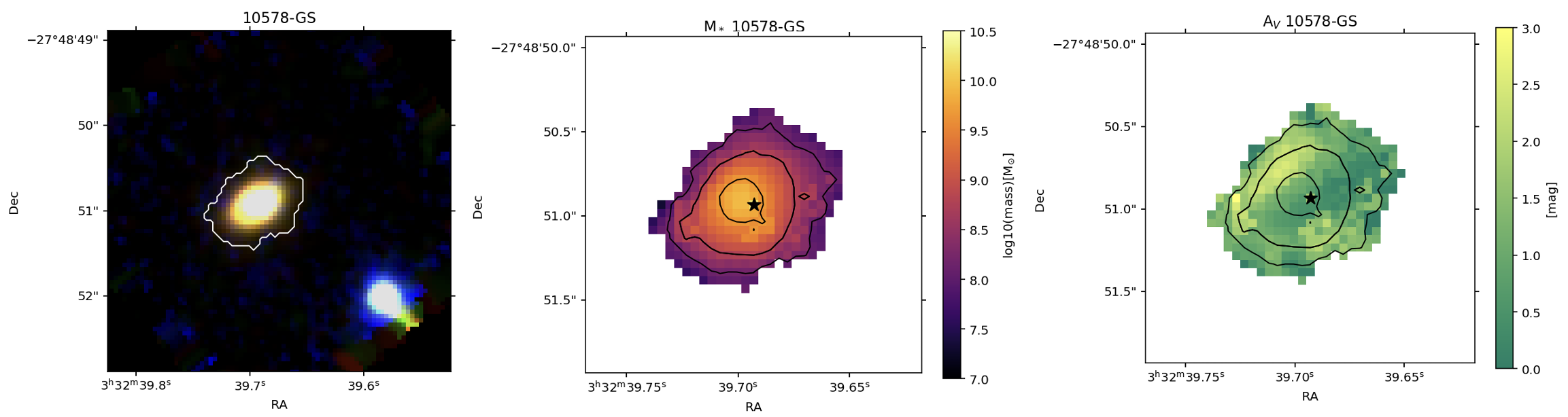}
    \caption{Resolved properties of GS10578. Left: three-colour image produced from rest-frame 2500\,\AA\ (blue), 5500\,\AA\ (green), and 7500\,\AA\ (red) bands, each with a bandwidth of 0.01\,$\mu$m. Centre: spatially resolved stellar mass map. Right: dust attenuation map expressed as $A_V$. White contour in the three colour image represents the  border of the aperture from where we extracted the unresolved spectra. Black contours represent arbitrary mass contours. Black stars represent the peak of the continuum at 5500~\AA. }
    \label{fig:10578-GS}
\end{figure*}

\begin{figure*}
    \centering
    \includegraphics[width=1\linewidth]{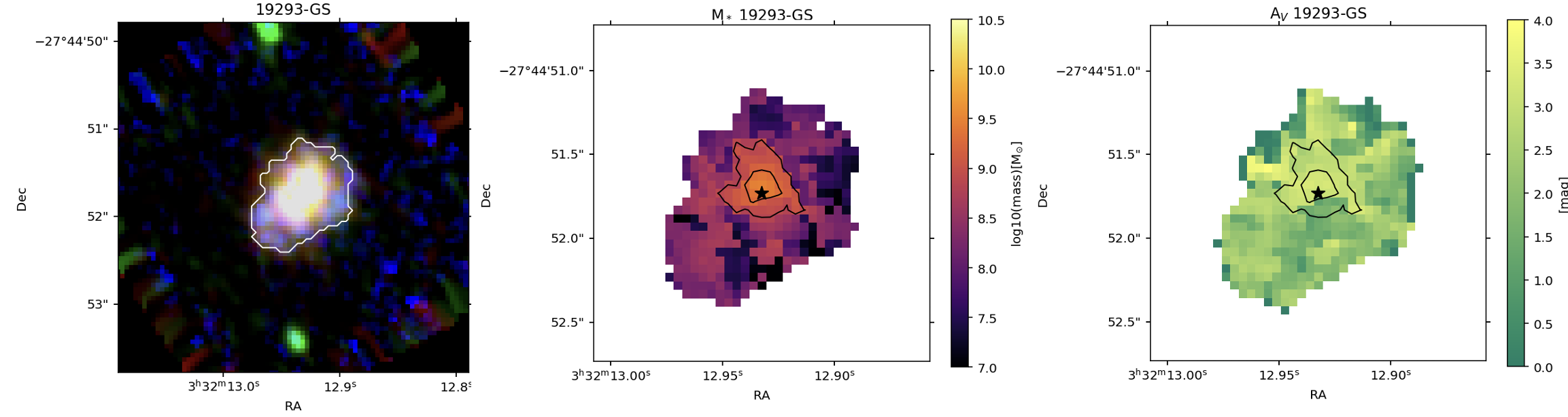}
    \caption{Same as Fig.~\ref{fig:10578-GS} for GS19293.}
    \label{fig:19293-GS}
\end{figure*}

\begin{figure*}
    \centering
    \includegraphics[width=1\linewidth]{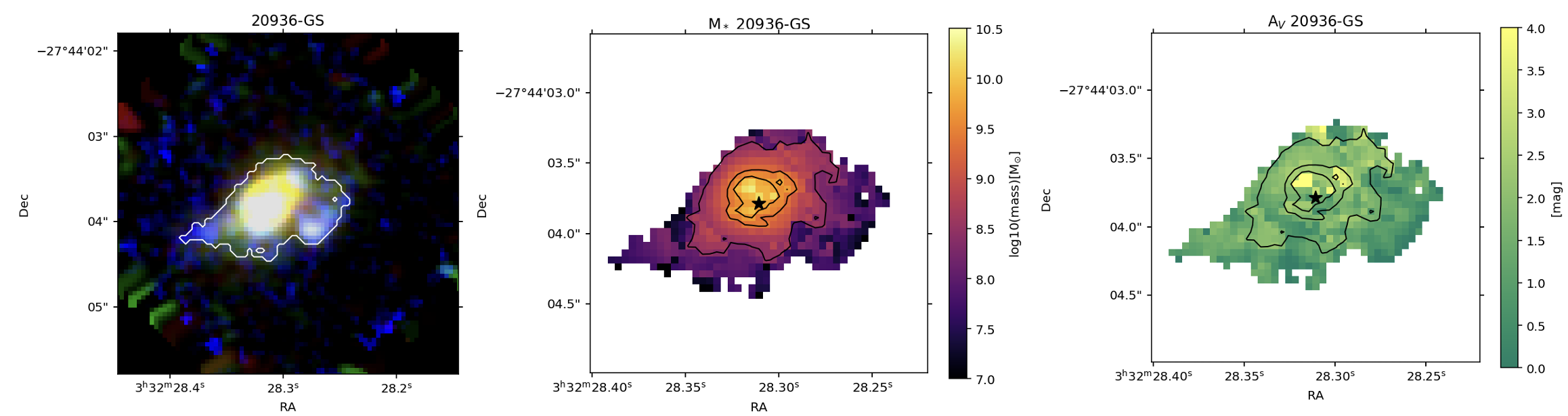}
    \caption{Same as Fig.~\ref{fig:10578-GS} for GS20936.}
    \label{fig:20936-GS}
\end{figure*}

\begin{figure*}
    \centering
    \includegraphics[width=1\linewidth]{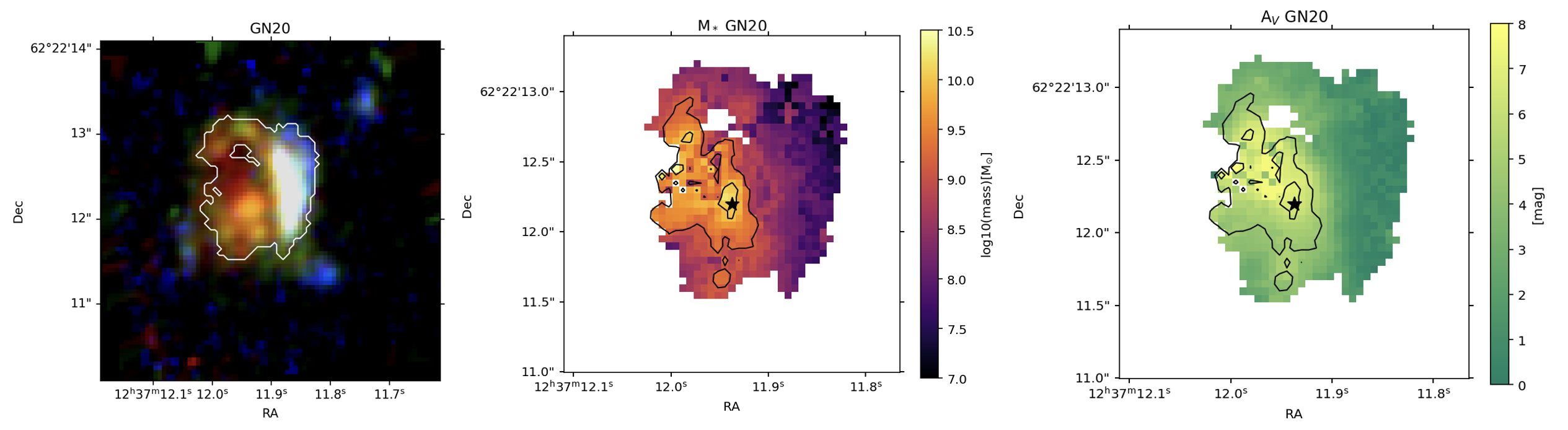}
    \caption{Same as Fig.~\ref{fig:10578-GS} for GN20.}
    \label{fig:GN20}
\end{figure*}

\begin{figure*}
    \centering
    \includegraphics[width=1\linewidth]{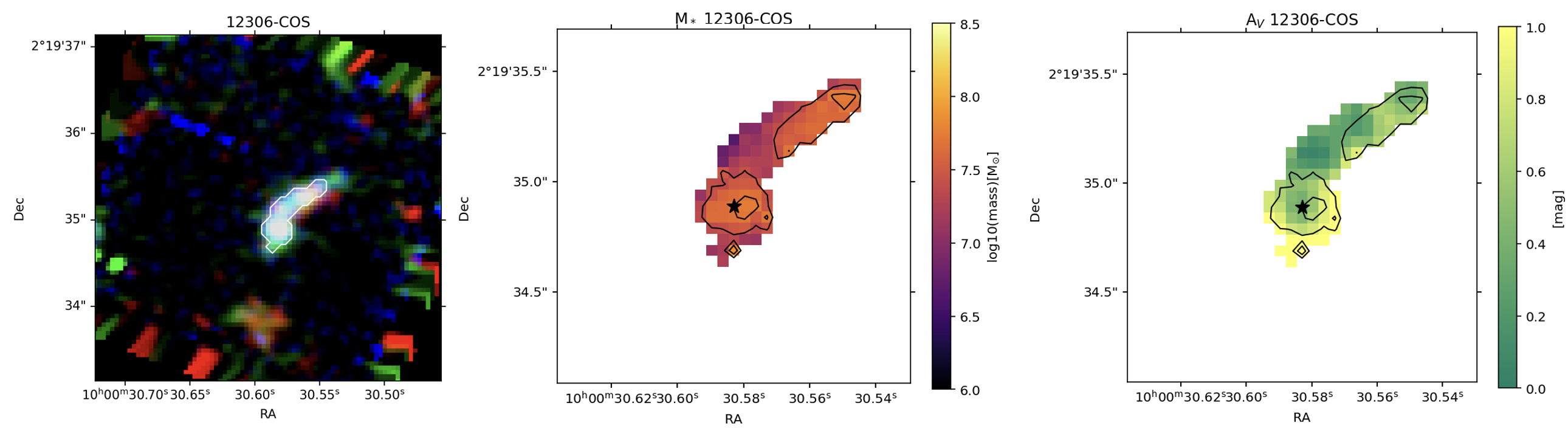}
    \caption{Same as Fig.~\ref{fig:10578-GS} for COS12306.}
    \label{fig:12306-COS}
\end{figure*}

Figures \ref{fig:10578-GS}-\ref{fig:12306-COS} represent the resolved stellar mass and the extinction maps of our targets derived from our fit together with the three colour image derived from NIRSpec \prism where the blue, green and red band represent the 2500A, 5500A and 7500A rest frame with a bandwidth of 0.01$\mu$m (2500A, 4000A and 5250A in case of EGSY8P7). In general, the continuum emission in sources hosting an AGN is concentrated in the nuclear region of the galaxy, while also showing extended features likely associated with blue star-forming clumps in a  disk or with merger-driven structures with lower mass compared to the central region. In the case of the two starburst galaxies, COS12306 and EGSYPz8, the stellar mass shows a more uniform spatial distribution, with a lower contrast between the inner and outer regions (\citealt{Tacchella2019}). We also note that in low-mass, high-redshift starbursts, the attenuation is lower compared to that in massive galaxies hosting an AGN (with an A$_\mathrm{V}$ lower than 1 in all regions). Moreover, the highest extinction values are not always coincident with the nucleus, as in the case of GS10578, where the highest extinction value lies in the NE-SW direction, where the outflow is directed (\citealt{Venturi2025}). 

The most extreme case of dust attenuation is GN20, which reaches an extinction of around 7 mag in the nuclear region. This is consistent with estimation of $A_V$ in \cite{Bik2024} through the Pa$\alpha$/Pa$\beta$ ratio, where they found $A_V > 5.6$ mag but it is smaller than $A_V = 17.2 \pm 0.4$ mag found comparing  the integrated SFR derived from Pa$\alpha$ flux, $\mathrm{SFR}_{\mathrm{Pa}\alpha} = 144 \pm 9\,M_{\odot}\,\mathrm{yr}^{-1}$, with the infrared-derived SFR (likely because of the presence of an AGN contributing to the infrared emission; e.g. \citealt{Duras2020}).

\begin{figure*}
    \centering
    \includegraphics[width=1\linewidth]{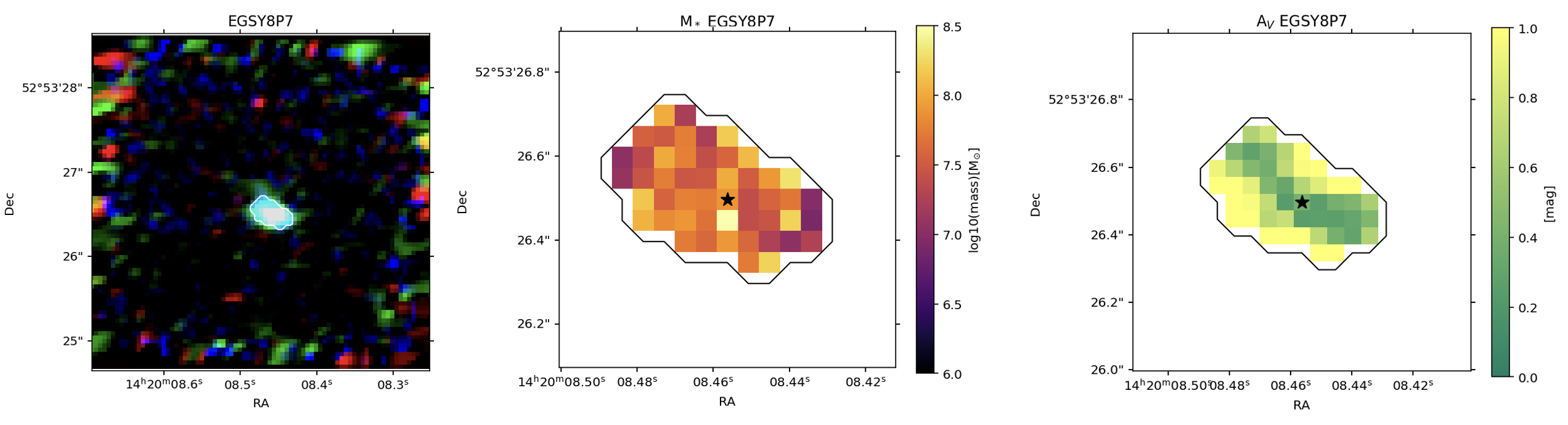}
    \caption{Same as Fig.~\ref{fig:10578-GS} for EGSY8p7.}
    \label{fig:EGSY8P7}
\end{figure*}



\section{Caveats on the interpretation of outflow impact}\label{app:Caveats}

The interpretation of the connection between outflows and star formation suppression presented in Sect. \ref{subsec:outflows} is subject to several limitations.

First, the mass loading factor $\eta$ is derived using only the ionized gas phase, while a potentially significant fraction of the outflowing material may reside in neutral or molecular phases. As a result, the values of $\eta$ reported in Table~\ref{tab:properties_outflow} should be considered lower limits to the total mass loading. This is supported by the case of GS10578, where \cite{DEugenio2024} found that the NaI mass outflow rate reaches $\sim 300$~\Msunyr, implying a $\eta \sim 4$.

Second, $\eta$ provides only a snapshot of the current balance between outflows and star formation. It does not capture the time variability of either quantity. Outflows may be episodic, and past feedback episodes, possibly stronger (or weaker) than those currently observed, may have contributed (or not) to shaping the SFHs. 

Third, values of $\eta\lesssim 1$ do not necessarily imply that outflows are inefficient in regulating star formation. Even relatively modest outflows can impact the gas reservoir by heating the ISM, disrupting dense gas, or preventing gas accretion, thereby influencing star formation on longer timescales (\citealt{Harrison2024}).

Fourth, the derivation of mass outflow rates depends on several assumptions, including geometry, electron density, and the spatial extent of the outflow. Different choices can lead to systematic uncertainties of up to an order of magnitude.

Finally, the comparison between outflow properties and SFHs is indirect. The SFHs trace star formation averaged over tens to hundreds of Myr, while outflows traced by emission lines reflect conditions on much shorter timescales  ($\sim 10$~Myr). This mismatch in timescales complicates a direct causal interpretation.
In Sect.~\ref{subsec:outflows} we assumed that if outflows are observed at the present epoch, it is plausible that similar episodes occurred in the past. For this reason, we preferred to compute $\eta$ as the ratio of $\dot{M}_{\mathrm{out}}$ to SFR$_{100}$; this long-term measurement is preferred to SFR$_{10}$ as we are more interested in the long-lasting effects of outflows, and not just the SFR variations possibly associated with instantaneous outflows traced by \OIII and \ha.

For these reasons, the results presented in Sect.~\ref{subsec:outflows} should be interpreted as indicative of possible connections between outflows and star formation regulation, rather than as definitive evidence of causality.

\section{GS20936}\label{sec:appE}
\begin{figure}
    \centering
    \includegraphics[width=0.99\linewidth]{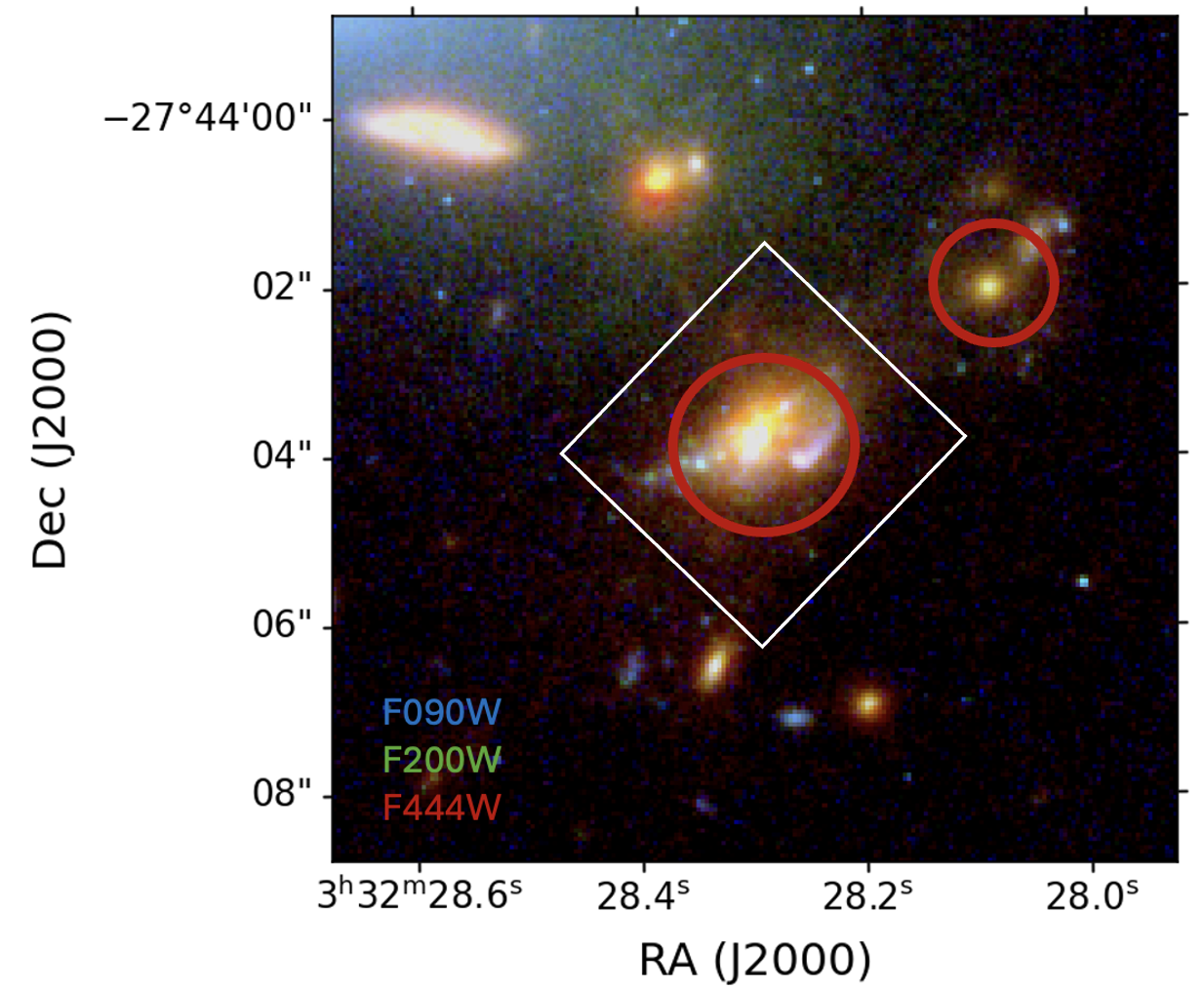}
    \caption{NIRCam image of GS20936 obtained by JADES (\citealt{Eisenstein2026}). The two red circles highlight the two main merging companion galaxies. The white contour represents the NIRSpec footprint.}
    \label{fig:rej20936}
\end{figure}

The NIRCam imaging of  obtained by JADES (\citealt{Eisenstein2026}) reveals an ongoing major merger, with prominent filamentary structures connecting the two galaxies within the region highlighted by the red circle in Fig.\ref{fig:rej20936}.

\section{Snapshot of progenitors of rejuvanated galaxies }\label{sec:Snapshot}

\begin{figure*}
    \centering
    \includegraphics[width=1\linewidth]{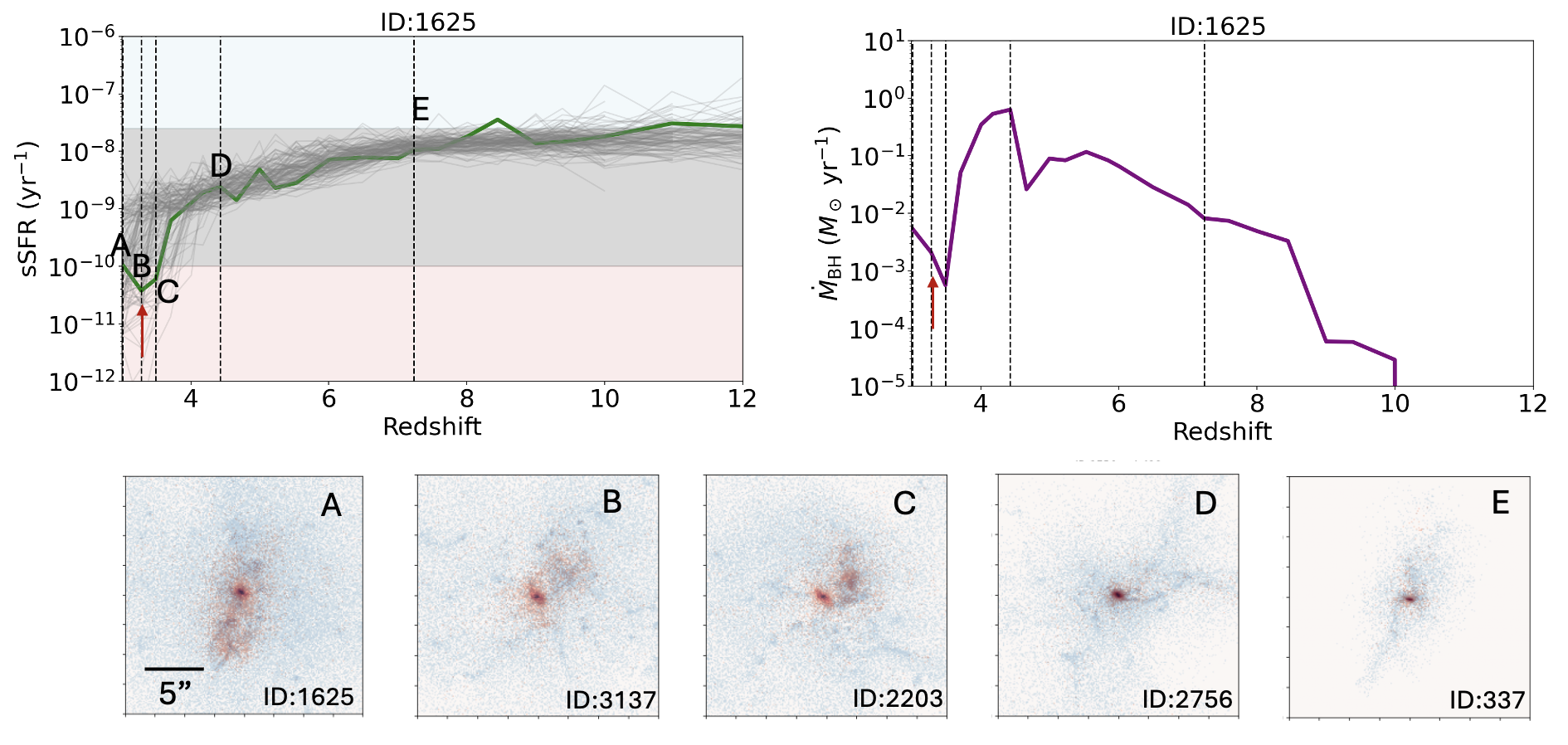}
    \caption{Upper panels: Left: sSFR of galaxy ID:1625 as a function of redshift (green curve), together with the sSFR histories of all 119 massive galaxies (grey curves). The red arrow marks the minimum of the sSFR along the main galaxy track. Right: black hole accretion rate as a function of redshift (purple curve). Lower panels: snapshots of the galaxy stellar mass (red) and gas mass (blue), shown at the redshifts corresponding to those in the upper panel. }
    \label{fig:ID1625}
\end{figure*}

\begin{figure*}
    \centering
    \includegraphics[width=1\linewidth]{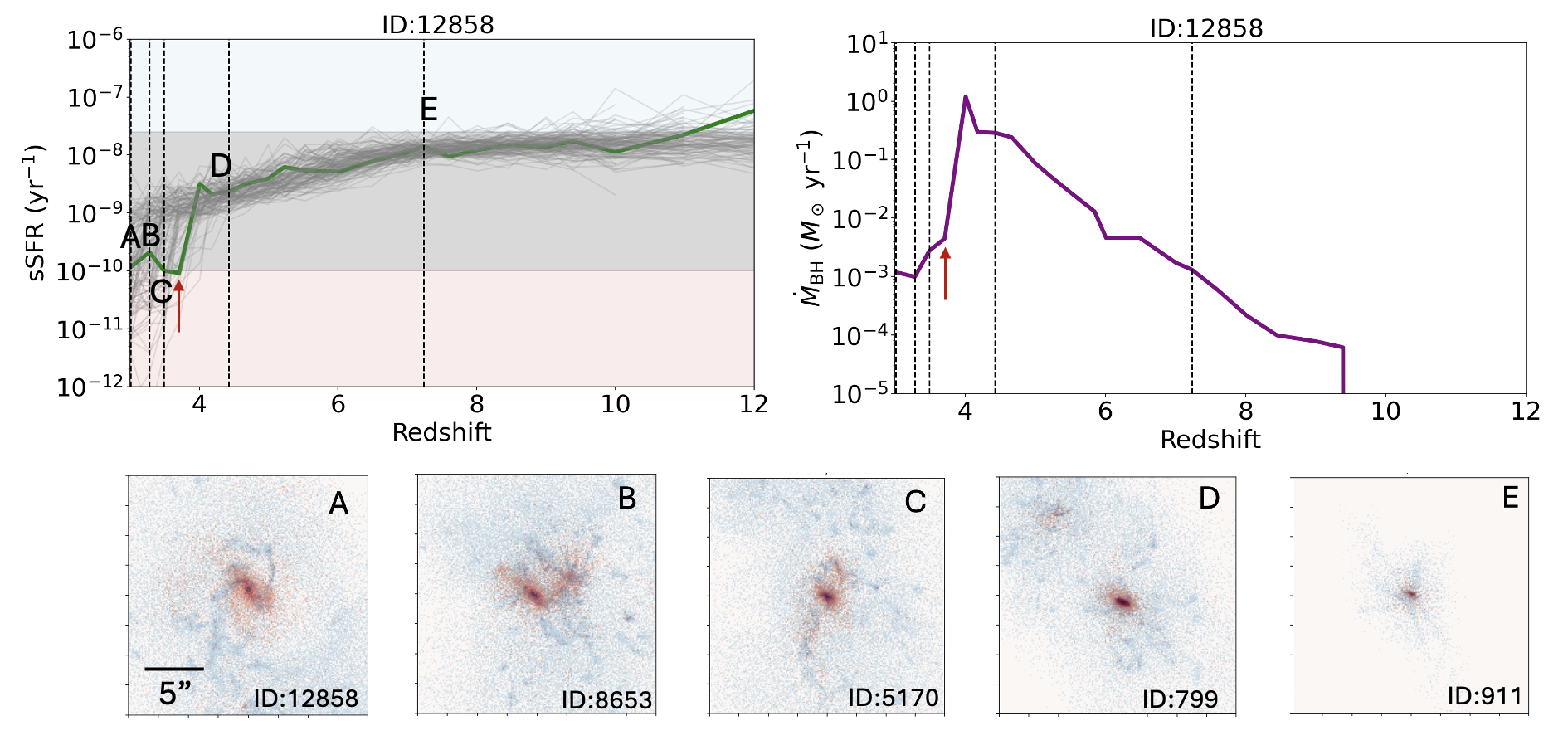}
    \caption{Snapshots of ID: 12858 at different redshift bin. Stellar mass in red and gas mass in blue. }
    \label{fig:ID12858}
\end{figure*}

\begin{figure*}
    \centering
    \includegraphics[width=1\linewidth]{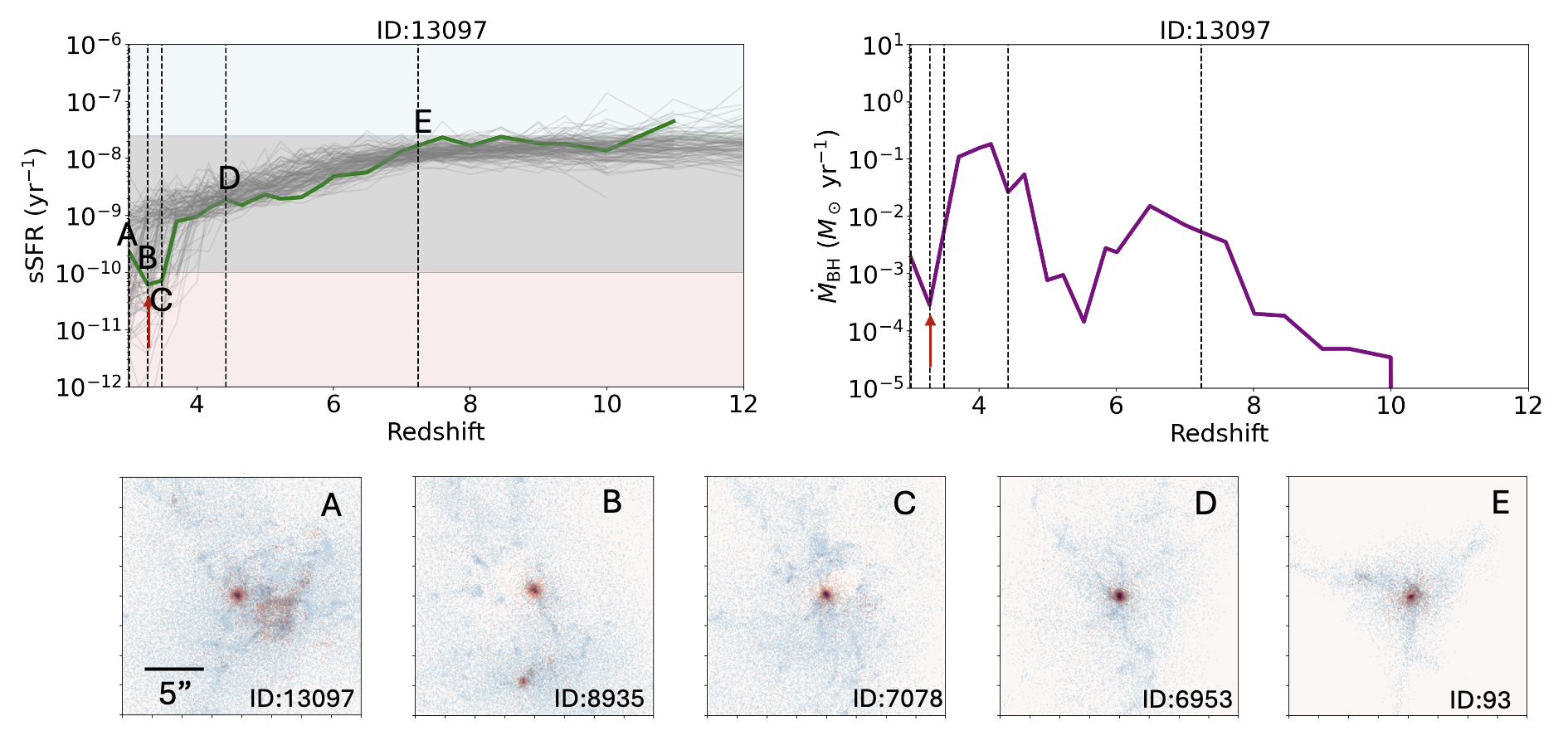}
    \caption{Same as Fig. for ID:13097.}
    \label{fig:ID13097}
\end{figure*}

\begin{figure*}
    \centering
    \includegraphics[width=1\linewidth]{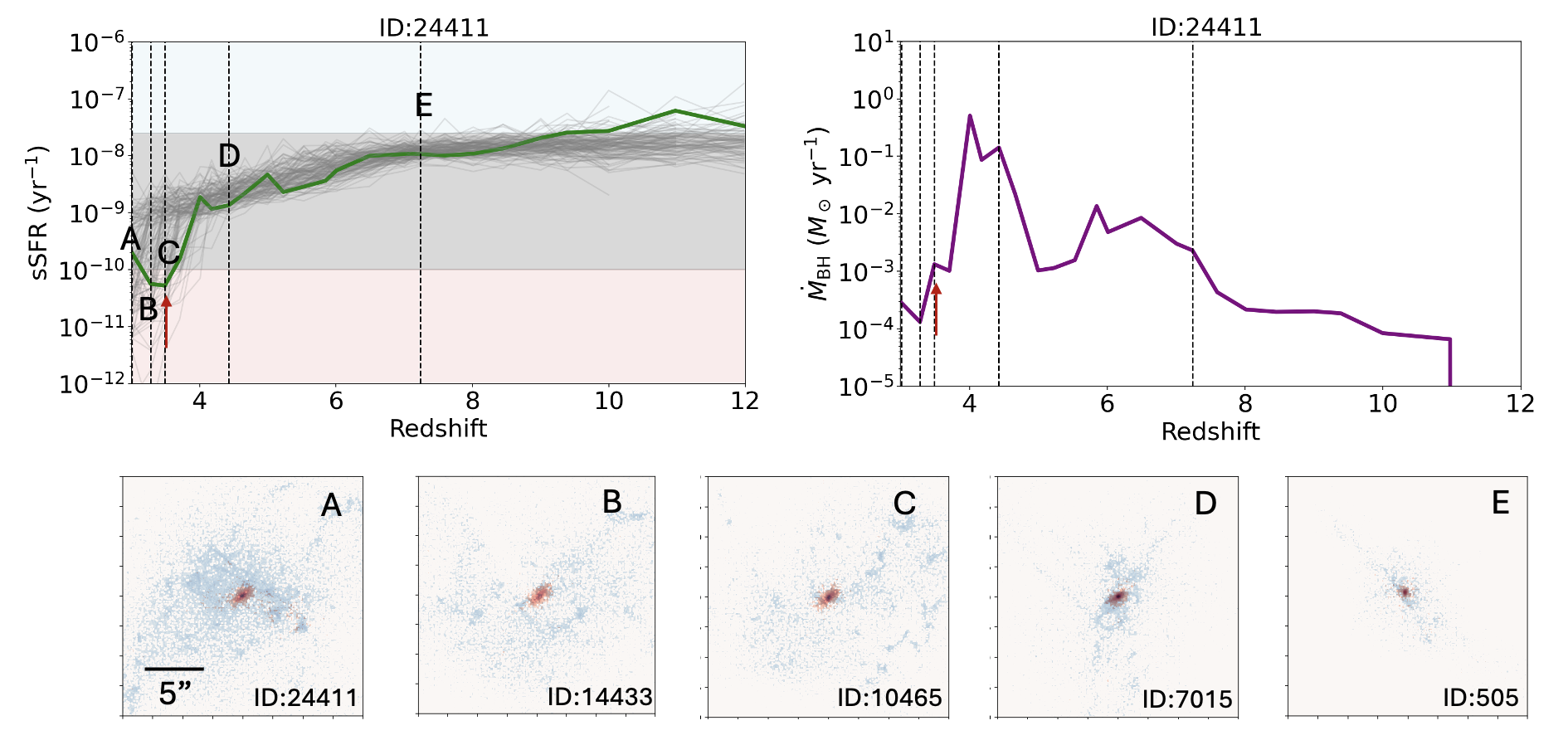}
    \caption{Same as Fig. for ID:24411.}
    \label{fig:ID24411}
\end{figure*}

\begin{figure*}
    \centering
    \includegraphics[width=1\linewidth]{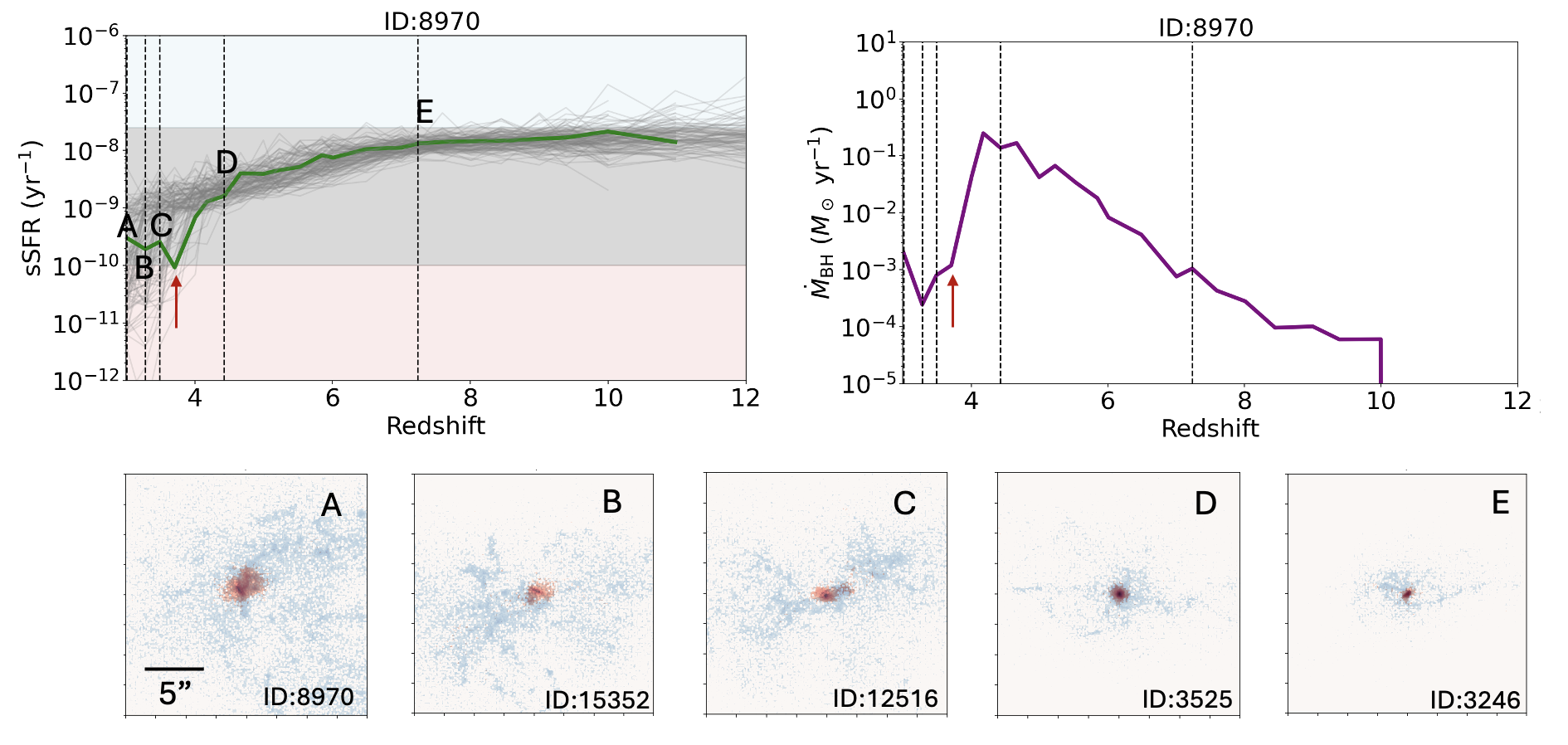}
    \caption{Same as Fig. for ID:8970. }
    \label{fig:ID8970}
\end{figure*}

\begin{figure*}
    \centering
    \includegraphics[width=1\linewidth]{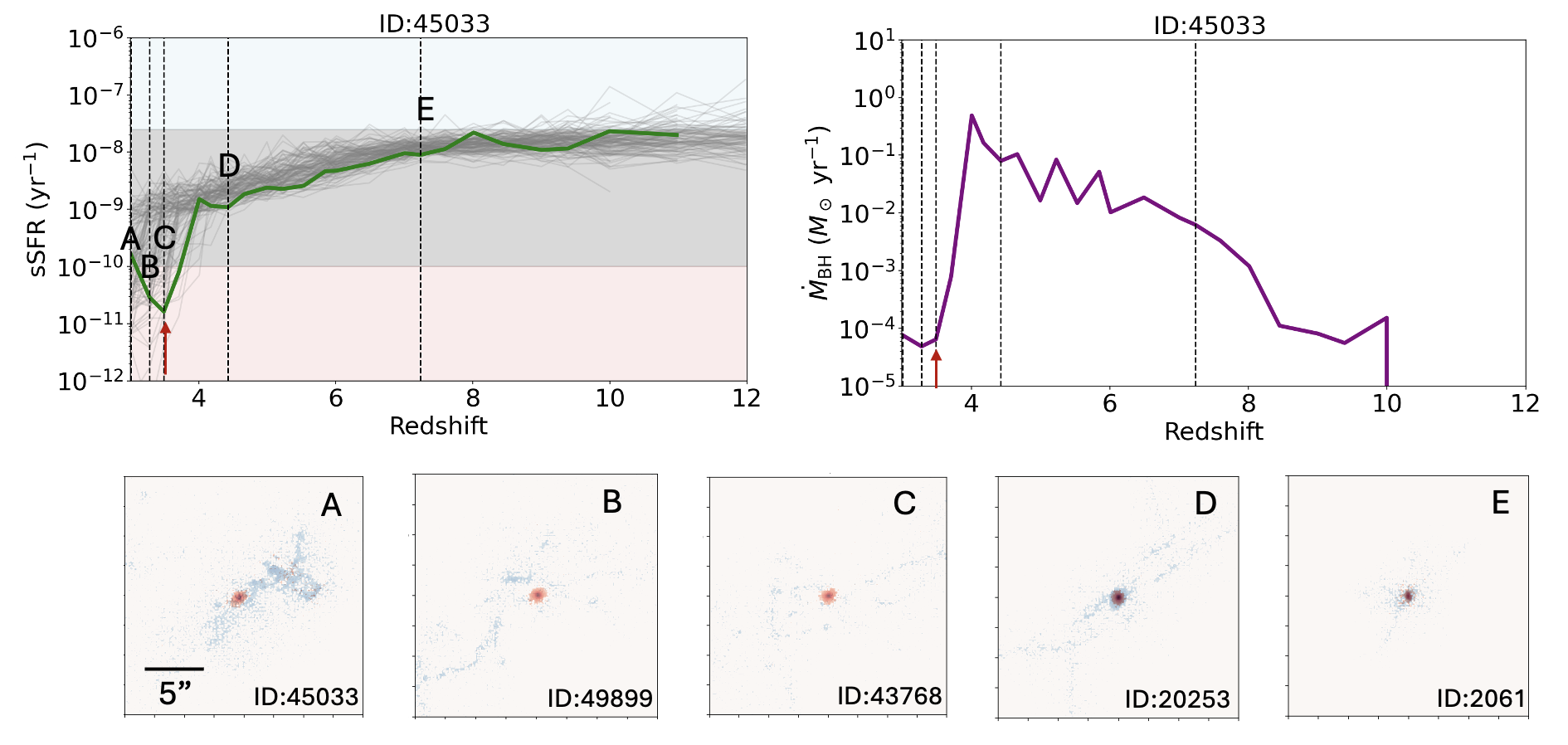}
    \caption{Same as Fig. for ID:45033. }
    \label{fig:ID45033}
\end{figure*}



\label{lastpage}
\end{document}